\documentclass[12pt]{article}
\usepackage{graphicx}
\usepackage{rotating}
\usepackage{amsmath}

\usepackage{epsfig,amsmath,amssymb,verbatim,mathrsfs}

\numberwithin{equation}{section}

\def\beq{\begin{eqnarray}}
\def\eeq{\end{eqnarray}}
\def\bea{\begin{eqnarray}}
\def\eea{\end{eqnarray}}

\newcommand{\gsim}{\lower.7ex\hbox{$\;\stackrel{\textstyle>}{\sim}\;$}}
\newcommand{\lsim}{\lower.7ex\hbox{$\;\stackrel{\textstyle<}{\sim}\;$}}

\def\stilde{\widetilde}

\newcommand{\newc}{\newcommand}
\newc{\Nc}{N_{c}}
\newc{\CG}{C_G}
\newc{\gp}{g'}
\newc{\stopi}{\stilde t_i}
\newc{\sboti}{\stilde b_i}
\newc{\staui}{\stilde \tau_i}
\newc{\stopj}{\stilde t_j}
\newc{\sbotj}{\stilde b_j}
\newc{\stauj}{\stilde \tau_j}
\newc{\stopI}{\stilde t_1}
\newc{\stopII}{\stilde t_2}
\newc{\sbotI}{\stilde b_1}
\newc{\sbotII}{\stilde b_2}
\newc{\stauI}{\stilde \tau_1}
\newc{\stauII}{\stilde \tau_2}
\newc{\sstop}{s_{t}}
\newc{\cstop}{c_{t}}
\newc{\ssbot}{s_{b}}
\newc{\csbot}{c_{b}}
\newc{\sstau}{s_{\tau}}
\newc{\cstau}{c_{\tau}}
\newc{\Sstop}{s_{2t}}
\newc{\Cstop}{c_{2t}}
\newc{\Ssbot}{s_{2b}}
\newc{\Csbot}{c_{2b}}
\newc{\Sstau}{s_{2\tau}}
\newc{\Cstau}{c_{2\tau}}
\newc{\salpha}{s_\alpha}
\newc{\calpha}{c_\alpha}
\newc{\Calpha}{c_{2\alpha}}
\newc{\Salpha}{s_{2\alpha}}

\newc{\sbetapm}{s_{\beta_\pm}}
\newc{\cbetapm}{c_{\beta_\pm}}
\newc{\Sbetapm}{s_{2 \beta_\pm}}
\newc{\Cbetapm}{c_{2 \beta_\pm}}
\newc{\sbetaO}{s_{\beta_0}}
\newc{\cbetaO}{c_{\beta_0}}
\newc{\SbetaO}{s_{2 \beta_0}}
\newc{\CbetaO}{c_{2 \beta_0}}
\newc{\vu}{v_u}
\newc{\vd}{v_d}
\newc{\seL}{\stilde e_L}
\newc{\smuL}{\stilde \mu_L}
\newc{\seR}{\stilde e_R}
\newc{\smuR}{\stilde \mu_R}
\newc{\suL}{\stilde u_L}
\newc{\sdL}{\stilde d_L}
\newc{\suR}{\stilde u_R}
\newc{\sdR}{\stilde d_R}
\newc{\scL}{\stilde c_L}
\newc{\ssL}{\stilde s_L}
\newc{\scR}{\stilde c_R}
\newc{\ssR}{\stilde s_R}
\newc{\snue}{\stilde \nu_e}
\newc{\snumu}{\stilde \nu_\mu}
\newc{\snutau}{\stilde \nu_\tau}
\newc{\Gpm}{G^\pm}
\newc{\Hpm}{H^\pm}
\newc{\FFbS}{\overline{FF}S}
\newc{\FFbV}{\overline{FF}V}
\newc{\FSS}{F_{SS}}
\newc{\FSSS}{F_{SSS}}
\newc{\FFFS}{F_{FFS}}
\newc{\FFFbS}{F_{\overline{FF}S}}
\newc{\FSSV}{F_{SSV}}
\newc{\FVS}{F_{VS}}
\newc{\FVVS}{F_{VVS}}
\newc{\FFFV}{F_{FFV}}
\newc{\FFFbV}{F_{\overline{FF}V}}
\newc{\Fgauge}{F_{\rm gauge}}
\newc{\DRbarprime}{$\overline{\rm DR}'$ }
\newc{\DRbar}{$\overline{\rm DR}$ }
\newc{\MSbar}{$\overline{\rm MS}$ }
\newc{\Yu}{{\bf Y}_u}
\newc{\Yd}{{\bf Y}_d}
\newc{\Ye}{{\bf Y}_e}
\newc{\Au}{{\bf a}_u}
\newc{\Ad}{{\bf a}_d}
\newc{\Ae}{{\bf a}_e}
\newc{\bm}{{\bf m}}
\newc{\zhol}{Z^{\rm hol}}

\newcommand{\mr}[1]{\mathrm{#1}}

\begin{document}

\setlength{\baselineskip}{0.2in}


\begin{titlepage}
\noindent
\begin{flushright}
MCTP-10-09\\
\end{flushright}
\vspace{1cm}

\begin{center}
  \begin{Large}
    \begin{bf}
Extracting the Dark Matter Mass from Single Stage Cascade Decays at the LHC \\
     \end{bf}
  \end{Large}
\end{center}
\vspace{0.2cm}

\begin{center}

\begin{large}
Timothy Cohen, Eric Kuflik, and Kathryn M. Zurek\\
\end{large}
\vspace{0.3cm}
  \begin{it}

Michigan Center for Theoretical Physics (MCTP)\\
Department of Physics, University of Michigan\\
Ann Arbor, Michigan 48109, USA\\
\vspace{0.5cm}
\end{it}

\end{center}

\center{\today}

\begin{abstract}
We explore a variant on the $M_{T2}$ kinematic variable which enables dark matter mass measurements for simple, one stage, cascade decays.  This will prove useful for constraining a subset of supersymmetric processes, or a class of leptophilic dark matter models at the LHC.  We investigate the statistical reach of these measurements and discuss which sources of error have the largest effects.  For example, we find that using only single stage cascade decays with initial state radiation, a measurement of a 150 GeV dark matter candidate can be made to $\mathcal{O}(10\%)$ for a parent mass of 300 GeV with a production cross section of 100 fb and $100 \mbox{ fb}^{-1}$ of integrated luminosity.
\end{abstract}

\vspace{1cm}

\end{titlepage}

\setcounter{page}{2}


\vfill\eject



\newpage

\section{Introduction}
The Large Hadron Collider (LHC) is operational, and taking data, with expected gains in energy and luminosity over the next few years.  One important mission for the LHC will be to create dark matter (DM) which appears as missing energy in the reconstructed event.  Following a significant missing energy observation, the challenge will be to measure the properties of the DM candidate with sufficient accuracy to compare against cosmological and astrophysical constraints, such as the observed DM relic abundance and direct and indirect detection experiments.  Thus, determining the mass of the DM particle 
will have tremendous ramifications for astrophysics and cosmology.

\begin{figure}[htp]
\label{fig:decaychain}
\centering
\includegraphics[width=0.5\textwidth]{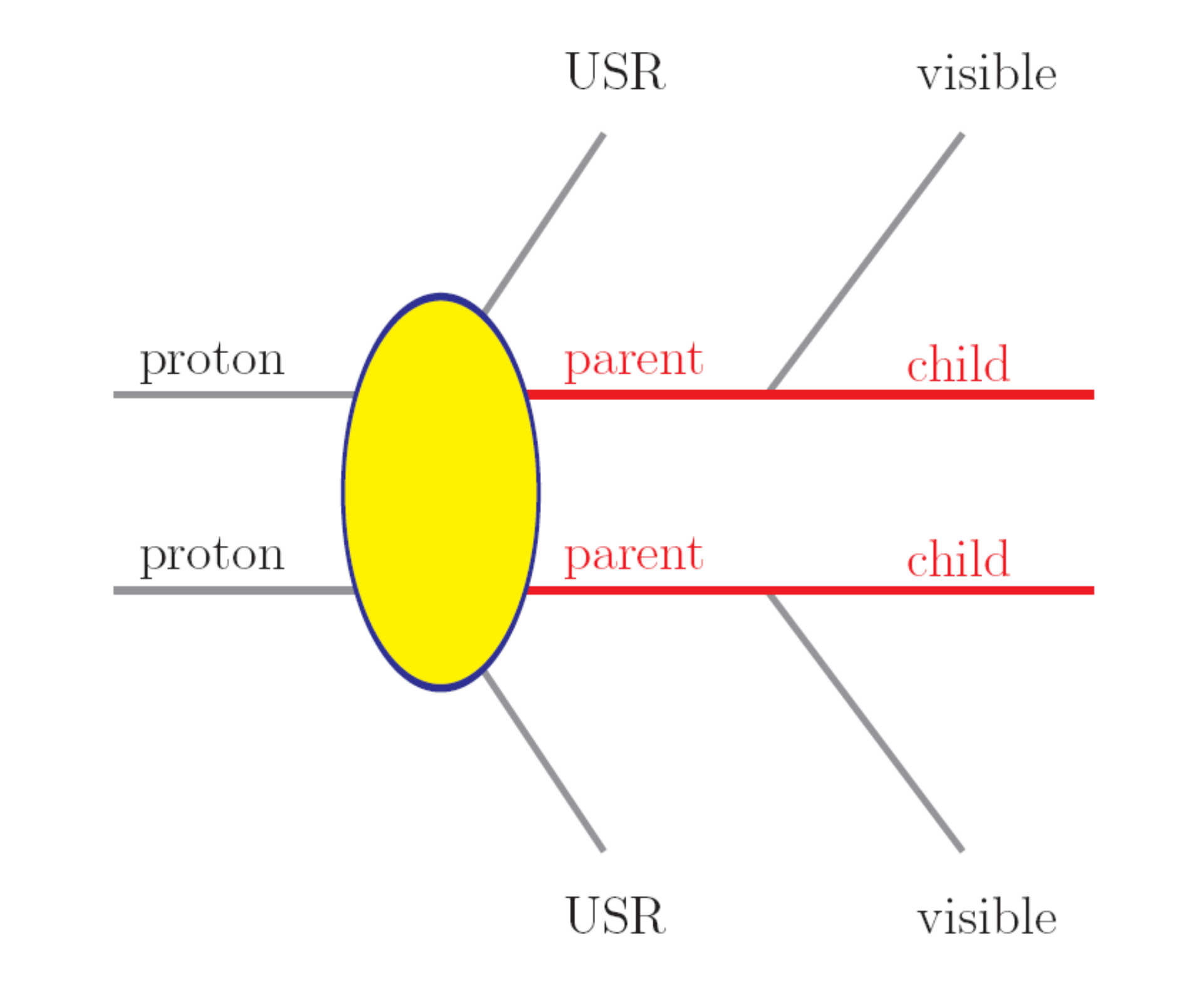}
\caption{ Schematic representation of the $n=1$ class of processes considered in this work, with additional Up-Stream Radiation (USR).  The parent particle is the state which decays to the visible particles and the child DM particles. }
\end{figure}

Making DM mass measurements at the LHC, for example in models of supersymmetry (SUSY) or Universal Extra Dimensions (UED), is a difficult problem, since the DM particle is typically produced in pairs as products of complicated decay chains of parent particles.  In fact, the number of states participating in the event can vary dramatically depending on the specific model.  The identities, couplings, and masses of the particles involved in these processes may be unknown. 
Let $n$ be the number of steps in the cascade between the production of the parent and the appearance of the DM child in the event.  For $n > 1$, if all visible particles in the decay are detected, all masses of the parent, intermediate and visible and invisible child particles can, in principle, be determined uniquely (see for example \cite{kong} for a discussion).  The simplest case of $n = 1$ proves to be more challenging.  In Fig.~\ref{fig:decaychain} we show a schematic of an $n=1$ process.  We have also included the possibility that additional visible states are produced before the parents, which we refer to as Up-Stream Radiation (USR).  In Sec.~\ref{sec:MT2Prelim} below, we will discuss the relevance of USR for DM mass determination.  Refs.~ \cite{kink2,USR} also study $n=1$ decay chains.

The motivation for studying DM mass determination in $n=1$ processes is many fold -- we mention two here.  First, within SUSY or UED, $n=1$ processes with additional USR can be important.  For example, decays $\tilde{\ell}^{\pm} \rightarrow \ell^{\pm}\, \tilde{\chi}^0$ with initial state radiation, and $\tilde{q} \rightarrow j\, \tilde{\chi}^{\pm} \rightarrow j\,\ell^{\pm} \tilde{\nu}$ (for a sneutrino lightest SUSY particle), are of the type shown in Fig.~\ref{fig:decaychain}, where $\tilde{\ell}^{\pm}$ is a slepton, $\ell^{\pm}$ is a lepton, $\tilde{\chi}^0$ is a neutralino, $\tilde{q}$ is a squark, $j$ is a jet, $\tilde{\chi}^{\pm}$ is a chargino and $\tilde{\nu}$ is a sneutrino.  Although higher $n$ chains may also be present in many models, the combinatoric backgrounds can make mass extraction in such decay chains complicated.  By contrast $n = 1$ events are clean, and involve only two visible objects plus missing transverse momentum (hereafter referred to as missing energy).  Also, since one will potentially observe $n=1$ chains if one of these theories is correct, it will be useful to extract as much information as possible from these signals.  Second, the observations of astrophysical anomalies, \emph{e.g.} PAMELA \cite{Adriani:2008zr} and Fermi \cite{Abdo:2009zk}, have led many to conjecture that the DM is leptophilic.  Models which generate such signals can, for example, be constructed by connecting the DM to the lepton asymmetry \cite{PAMELAADM}, or by positing that mixed sneutrinos constitute the DM \cite{mixedsneutrino}.  The simplest such dark sectors involve only a new mediator state and the leptophilic DM state, so that the DM is produced at a collider through the leptonic decay of the mediator.  Hence, the study of these processes is well motivated.  The reader is referred to Appendix \ref{sec:Models} for more detail on models where $n=1$ decay chains with USR are important.

As shown in Appendix \ref{sec:PhaseSpaceAndMT2}, the phase space for $n=1$ processes without USR depends on the combination $\mu = (m_p^2-m_c^2)/(2\,m_p)$ and weakly on $\hat{s}/(4\, m_p^2)$ where $m_p$ is the parent mass, $m_c$ is the child mass and $\sqrt{\hat{s}}$ is the partonic center-of-mass energy.  Hence, extracting $\mu$ is simple, while measuring $m_p$ proves to be more challenging.  Current experimental methods for mass determination in events with missing energy rely on matrix element techniques.  Here, one begins by assuming a model which implies a matrix element with additional dependence on $m_p$.  Then by fitting measured differential distributions, one can extract, in addition to the combination $\mu$, the overall mass scale $m_p$ by observing how quickly the event rate falls off with $\sqrt{\hat{s}}$.

In this paper we explore a different technique where the overall mass scale is determined from the transverse boosts given to the parent particles by USR.  Since the boost depends only on $m_p$, \emph{i.e.} it is independent of the matrix element, the result is a model independent method for determining the overall mass scale.  We explore a particular $M_{T2}$ variant proposed in \cite{bowls}, which utilizes events with USR to separately extract the parent and child masses.  We carry out the first full scale simulation of these $M_{T2}$ based variants for dark matter mass determination, including detector effects, emphasizing the size of statistical errors and discussing various difficulties this method presents.  

The outline of this paper is as follows.  We begin with a discussion of the $M_{T2}$ variable, and the possibility of extracting parent and child mass separately in $n = 1$ events with USR.  Next we turn to a numerical analysis of this $M_{T2}$ based method and its efficiency in DM mass determination for a given number of $n=1$ events at the LHC.  We then discuss additional sources of error beyond those explicitly contained in the previous section.  Finally, we conclude.  In Appendix A we outline some example models where this method would be relevant and in Appendix  B we show how the phase space for $n=1$ processes depends on the $M_{T2}$ endpoint.

\section{$M_{T2}$ Preliminaries}\label{sec:MT2Prelim}
We begin by reviewing the $M_{T2}$ variable \cite{MT2}.  Since the LHC is a hadron collider, the initial parton longitudinal momenta are unknown.  Hence, only the total transverse momentum is constrained to be zero, and thus it becomes necessary to use transverse variables, such as $M_{T2}$, a generalization of the transverse mass (see also \cite{generalization,IWKim}).  For the class of processes studied here (see Fig.~\ref{fig:decaychain}), there will be two missing particles in each event, so that the 4-momenta of the invisible child particles cannot be determined.  Thus, only the total transverse missing momentum, $\vec{p}_T^\mr{\,\,\,miss}$, can be measured.  In addition, the child particle mass, $m_c$, is not known.  However, a trial DM mass can be guessed, $\tilde{m}_c$, and $M_{T2}$ formed for each event as 
\begin{equation}\label{eq:MT2}
M_{T2}(\tilde{m}_c)  \equiv \mbox{min}\left[ \mbox{max}\left\{M_T^{(1)},M_T^{(2)} \right\} \right],
\end{equation}
where the minimization is performed over trial missing momenta for the two child particles, $\vec{p}_T^\mr{\,\,\,miss(1)} $, $\vec{p}_T^\mr{\,\,\,miss(2)}$, subject to the constraint that their sum be the total missing $\vec{p}_T$:
\begin{equation}
\vec{p}_T^\mr{\,\,\,miss(1)} + \vec{p}_T^\mr{\,\,\,miss(2)} = \vec{p}_T^\mr{\,\,\,miss} =  - \vec{p}_T^{\mr{\,\,\,vis}(1)} - \vec{p}_T^{\mr{\,\,\,vis}(2)},
\label{pTbalance}
\end{equation}
where $\vec{p}_T^{\mr{\,\,\,vis}(i)}$ is the transverse momentum of the $i^\mr{th}$ visible particle, and we are neglecting here the possibility of additional USR.  In Eq.~(\ref{eq:MT2}), $M_T^{(i)}$ is the transverse mass of the visible and child particles using the guessed missing momentum, $\vec{p}_T^{\mr{\,\,\,miss}(i)} $, and child trial mass $\tilde{m}_c$:
\begin{equation}
M_T^{(i)} = \sqrt{\left({m_{\mr{vis}}^{(i)}}\right)^2+ \tilde{m}_c^2 + 2 \left(E_T^{\mr{vis}(i)}E_T^{\mr{miss}(i)} - \vec{p}_T^{\mr{\,\,\,vis}(i)} \cdot \vec{p}_T^{\mr{\,\,\,miss}(i)} \right)},
\end{equation}
where $m_{\mr{vis}}^{(i)}$ is the mass of the $i^\mr{th}$ visible particle.  The energies are formed in the usual way,
\begin{equation}
E_T^\mr{{vis}(i)} \equiv \sqrt{\left({\vec{p}_T^{\mr{\,\,\,vis}(i)}}\right)^2 +\left(m_{\mr{vis}}^{(i)}\right)^2},~~~E_T^{\mr{miss}(i)} \equiv \sqrt{\left({\vec{p}_T^{\mr{\,\,\,miss}(i)}}\right)^2 +{\tilde{m}_c}^2 }.
\end{equation}
When there is no USR, there exists a value of $M_{T2}$, referred to as an endpoint, above which the differential cross section, $\mr{d}\sigma/\mr{d}M_{T2}$, rapidly approaches zero, which is given by \cite{Cho:2007dh}
\begin{equation}\label{eq:MT2Max}
M_{T2}^\mr{max} = \mu + \sqrt{\mu^2+\tilde{m}_c^2}, 
\end{equation}
where
\begin{equation}\label{eq:muDef}
\mu \equiv \dfrac{m_p^2-m_c^2}{2\, m_p},
\end{equation}
is the momentum of the invisible child in the parents' rest frame.  As we show in Appendix \ref{sec:PhaseSpaceAndMT2}, $n = 1$ chains only depend on $\mu$ up to small corrections due to the parent mass (which is exploited by the matrix element methods).  Methods which do not capitalize on these corrections do not have enough information to extract both masses separately.  
This neglects, however, the potential for additional USR in the event. The USR can be in the form of jets coming from the initial state QCD radiation (ISR), or jets coming from the decays of heavy colored objects in $n>1$ processes, where the decay chain ends in the $n=1$ process of interest. By including the USR, Eq. (\ref{pTbalance}) no longer obtains, and instead the total momentum of the visible and invisible particles must be balanced against the momentum of the radiation, $\vec{p}_T^\mr{\,\,\,USR} \equiv \vec{P}_T$,
\begin{equation}
\vec{p}_T^\mr{\,\,\,miss(1)} + \vec{p}_T^\mr{\,\,\,miss(2)} = \vec{p}_T^\mr{\,\,\,miss} =  - \vec{p}_T^\mr{\,\,\,vis(1)} - \vec{p}_T^\mr{\,\,\,vis(2)} - \vec{P}_T. 
\label{pTbalanceUSR}
\end{equation}

Now the $M_{T2}$ endpoint will depend on the upstream momentum \cite{kong,USR}: 
\begin{equation}
M_{T2}^\mr{max}(\tilde{m}_c,P_T) = \left\{
\begin{array}{cc}
\left[\left(\mu(P_T) + \sqrt{\left(\mu(P_T)+\frac{P_T}{2}\right)^2 + \tilde{m}_c^2}\right)^2 - \frac{P_T^2}{4} \right]^{1/2}, &  \mbox{if } \tilde{m}_c \leq m_{c} \\ 
\left[\left(\mu(-P_T) + \sqrt{\left(\mu(-P_T)-\frac{P_T}{2}\right)^2 + \tilde{m}_c^2}\right)^2 - \frac{P_T^2}{4} \right]^{1/2}, &  \mbox{if } \tilde{m}_c \geq m_{c} 
\end{array}
\right.
\label{eq:MT2withUSR}
\end{equation}
 and
\begin{equation}
\mu(P_T) \equiv \dfrac{m_p^2-m_c^2}{2\, m_p}\left( \sqrt{1+\left(\frac{P_T}{2\, m_p}\right)^2}-\frac{P_T}{2\, m_p} \right).
\label{kineqn}
\end{equation}
The functional form for the $M_{T2}$ endpoint depends on whether the test mass is larger or smaller than the true DM mass.  Hence, there is a discontinuity in the derivative with respect to the trial child mass of Eq.~(\ref{eq:MT2withUSR}) above and below the true DM mass, $m_c$, giving rise to a kink \cite{kink2,kink} in the $M_{T2}^\mr{max}(\tilde{m}_c,P_T)$ curve which can be utilized for extracting additional information beyond Eq.~(\ref{eq:MT2Max}).  In principle, given an event with a specific value for the $P_T$ of the USR, one can now extract the parent and child masses.  However, since one must do this analysis for a particular bin in $P_T$, there is competition between the size of the bin -- small bins imply small statistical samples -- and the accuracy of the measurement.  

\begin{figure}[htp]
\centering
\includegraphics[width=0.8\textwidth]{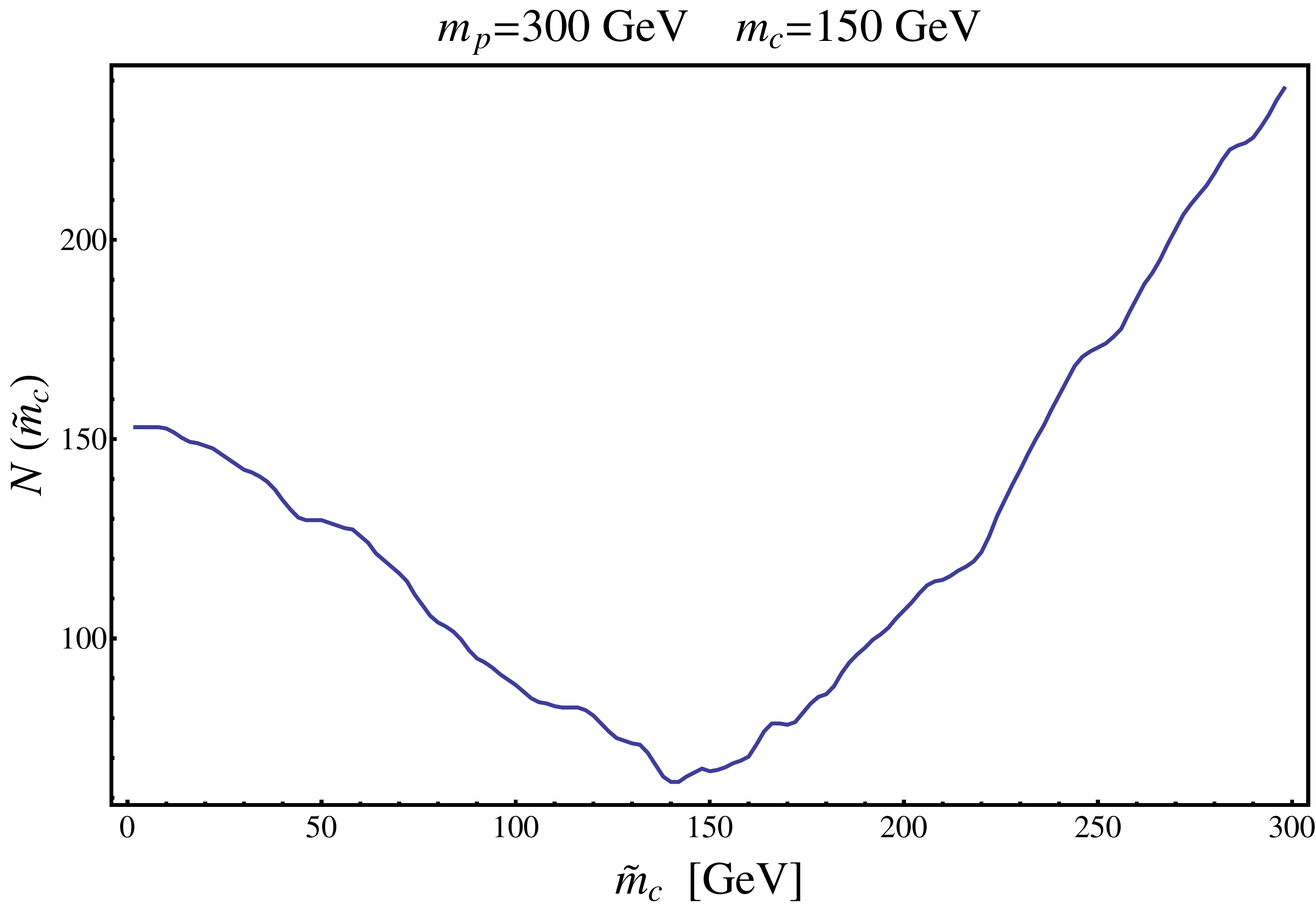}
\caption{  Example of an $M_{T2}$ bowl for 50,000 smuon pair production events with QCD USR.  The parent mass is 300 GeV and the child mass is 150 GeV.  The events were run through the PGS detector simulator.}
\label{fig:exampleBowl}
\end{figure}

Another method was proposed in \cite{bowls}, which sidesteps the problem of binning by utilizing the whole range of $P_T$.  From Eqs.~(\ref{eq:MT2Max}) and (\ref{eq:MT2withUSR}), it can be seen that $M_{T2}^\mr{max}$ is unchanged by the effects of the $P_T$ when $\tilde{m}_c = m_c$.  Furthermore, it has been shown \cite{bowls} that
\begin{equation}
M_{T2}^\mr{max}(\tilde{m}_c,P_T)- M_{T2}^\mr{max}(\tilde{m}_c,0)\geq 0,
\end{equation}
where the equality only holds when $\tilde{m}_c = m_c$.  Thus one can construct a new variable \cite{bowls}
\begin{equation}
N(\tilde{m}_c) \equiv \sum_{\mbox{all events}} \Theta\left(M_{T2}^\mr{measured}(\tilde{m}_c) - M_{T2}^\mr{max}(\tilde{m}_c,0)\right),
\label{eq:bowl}
\end{equation}
where $\Theta(...)$ is the Heaviside function and $M_{T2}^\mr{measured}(\tilde{m}_c)$ is the measured value of $M_{T2}(\tilde{m}_c)$.  It is this variable we will be minimizing to find the correct child mass.  In Fig.~\ref{fig:exampleBowl}, we plot $N(\tilde{m}_c)$ vs. $\tilde{m}_c$ for $m_p = 300 \mbox{ GeV}$ and $m_c = 150\mbox{ GeV}$.  Since the shape is ``bowl"-like, we refer to this construction as an $M_{T2}$ bowl.  Unless otherwise specified, all events were simulated with the MadGraph 4.4 event generator \cite{Alwall:2007st}, showered by PYTHIA 6.4 \cite{Sjostrand:2006za}, and run through the detector simulation software PGS 3.3 \cite{PGS}.  Note that we use the MadGraph default settings which defines a lepton as having $p_T > 10$ GeV and a jet as having $p_T > 20$ GeV.

\section{Mass Determination from $M_{T2}$ Bowls}
In this section we will calculate the statistical errors for child mass determination with $M_{T2}$ bowls.  Clearly, Eq.~(\ref{kineqn}) only depends on the kinematics of the event, \emph{i.e.} it is independent of the quantum numbers, including the spin, of the underlying particles.  Then, up to small corrections due to the steepness of the $M_{T2}$ distribution about this endpoint, there are only $\mathcal{O}(1)$ differences in the bowls around the minimum for different parent spins.  Hence, we can study the effectiveness of this variable for a wide variety of models by only scanning over the masses of the parent and child particles.  We take 
\begin{eqnarray}
m_p = 100 \mbox{ GeV}, &~~~~ & m_c = 25,~50,~75 \mbox{ GeV}, \nonumber \\
m_p = 300 \mbox{ GeV}, & ~~~~ &m_c = 75,~150,~225\mbox{ GeV}, \nonumber \\ 
m_p = 500 \mbox{ GeV}, & ~~~~ & m_c = 125,~250,~375 \mbox{ GeV} 
\end{eqnarray}
as our benchmark parameters.

For reference we provide the overall cross section for these benchmark models in Table~\ref{tab:BenchmarkCrosssections}, where we have assumed that the production occurs via electroweak processes, including the effects of QCD ISR.  Neglecting diagrams which involve additional new-physics states, the overall rates only depend on the spin of the parent up to $\mathcal{O}(1)$ factors due to the choice of $SU(2)\times U(1)$ representation.  For reference, the scalar example process is $p\,p\rightarrow \tilde{\ell}^+\,\tilde{\ell}^-\rightarrow \ell^+\,\ell^-\,\tilde{\chi}^0\,\tilde{\chi}^0$ where $\tilde{\ell}^{\pm}$ is a slepton and $\tilde{\chi}^0$ is the lightest neutralino.  This is the process we simulate for our benchmarks with QCD ISR.  For reference, a fermionic example process is $p\,p\rightarrow \tilde{\chi}^+\,\tilde{\chi}^-\rightarrow \ell^+\,\ell^-\,\tilde{\nu}\,\tilde{\nu}^*$ where $\tilde{\chi}^{\pm}$ is a chargino and $\tilde{\nu}$ is a sterile sneutrino.  For some details of these and other models which have $n=1$ processes, see Appendix \ref{sec:Models}.  Our results below will be given in terms of the number of events before cuts, so Table~\ref{tab:BenchmarkCrosssections} can be used to estimate the reach of actual models.

\begin{table}[h!]
\centering
\begin{tabular}{||c|c|c||}
\hline
\hline
  $m_p$  & $\sigma_\mr{scalar}$ & $\sigma_\mr{fermion}$ \\
\hline
 100 GeV  & 0.4 pb               & 20 pb \\
 300 GeV  & $9\times 10^{-3}$ pb & 0.4 pb \\
 500 GeV  & $10^{-3}$ pb         & $6\times 10^{-2}$ pb\\
\hline
\hline
\end{tabular}
\caption{Cross sections for electroweak pair production of parent particles with various masses and spins including the effects of QCD ISR.  We neglect any $t$-channel processes involving additional states.}
\label{tab:BenchmarkCrosssections}
\end{table}

There are also models which have more complicated decay chains but can be interpreted as $n=1$ processes with additional USR.  For example, one can have new colored objects which decay to jets and the parent particle.  As long as the USR can be distinguished from the decay product of the parent, our method is applicable.  This will improve the prospects for this method dramatically since the overall rate will increase due to colored production instead of electroweak production, and additionally the majority of events will have very hard $P_T$ for the USR.  Hence, we also choose a set of benchmark models with colored objects up-stream with masses
\begin{eqnarray}
m_\mr{col}  = 600 \mbox{ GeV},~~~  m_p = 300 \mbox{ GeV}, &~~ & m_c = 150 \mbox{ GeV}, \nonumber \\
m_\mr{col} = 1000 \mbox{ GeV},~~~  m_p = 300 \mbox{ GeV}, & ~~ &m_c = 150\mbox{ GeV}, \nonumber \\ 
m_\mr{col} = 1400 \mbox{ GeV},~~~m_p = 300 \mbox{ GeV}, & ~~ & m_c = 150 \mbox{ GeV},  
\end{eqnarray}
where $m_\mr{col}$ is the mass of the colored state which decays to the parent particle and jets.  The example process we will simulate for these benchmarks is $p\,p\rightarrow \tilde{q}\,\tilde{q}\rightarrow j\,j\,\tilde{\chi}^+\,\tilde{\chi}^-\rightarrow \ell^+\,\ell^-\,\tilde{\nu}\,\tilde{\nu}^*$ where $\tilde{q}$ is a squark, $\tilde{\chi}^{\pm}$ is a chargino and $\tilde{\nu}$ is a sterile sneutrino.  While one might be able to use additional handles from viewing such events as $n=2$ processes (instead of $n=1$ with USR), here we wish to examine only the effect of harder USR from colored particle decay on the error for DM mass determination in $n=1$.  For additional models of $n=1$ processes which can be produced in the decays of colored states see Appendix \ref{sec:Models}.   

\begin{figure}[htp]
\begin{center}
\includegraphics[width=0.8\textwidth]{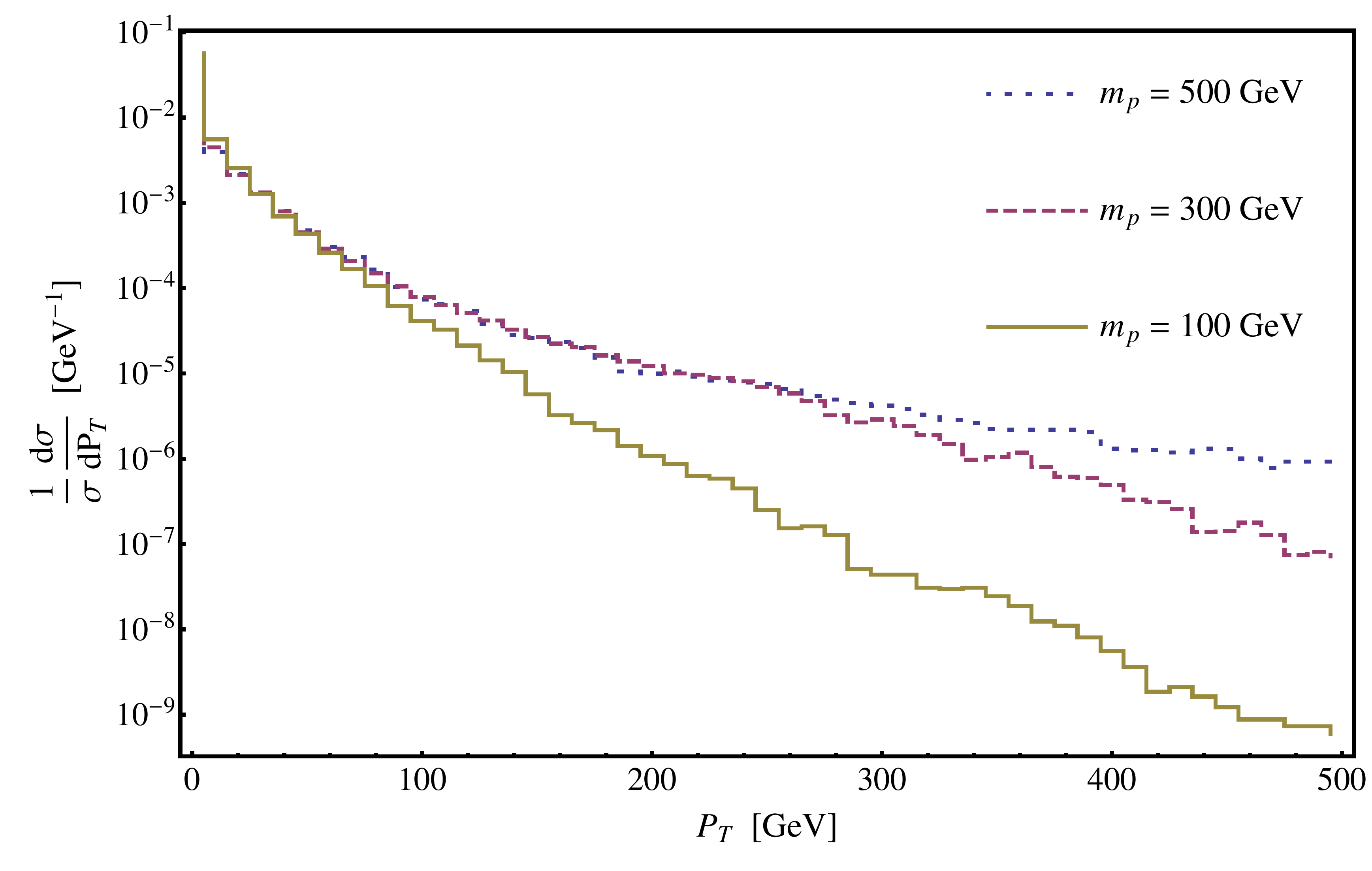}
\caption{  $P_T$ of the hardest jet for slepton events with only QCD ISR.  The blue dotted line is for $m_p = 500 \mbox{ GeV}$, the red dashed line is for $m_p = 300 \mbox{ GeV}$ and the yellow solid line is for $m_p = 100 \mbox{ GeV}$.  Note from Eq.~(\ref{eq:MT2withUSR}) that the correction to $M_{T2}$ due to USR is of the form $P_T/m_p$.}
\label{fig:pTelectroweak}
\end{center}
\end{figure}

\begin{figure}[htp]
\centering
\includegraphics[width=0.8\textwidth]{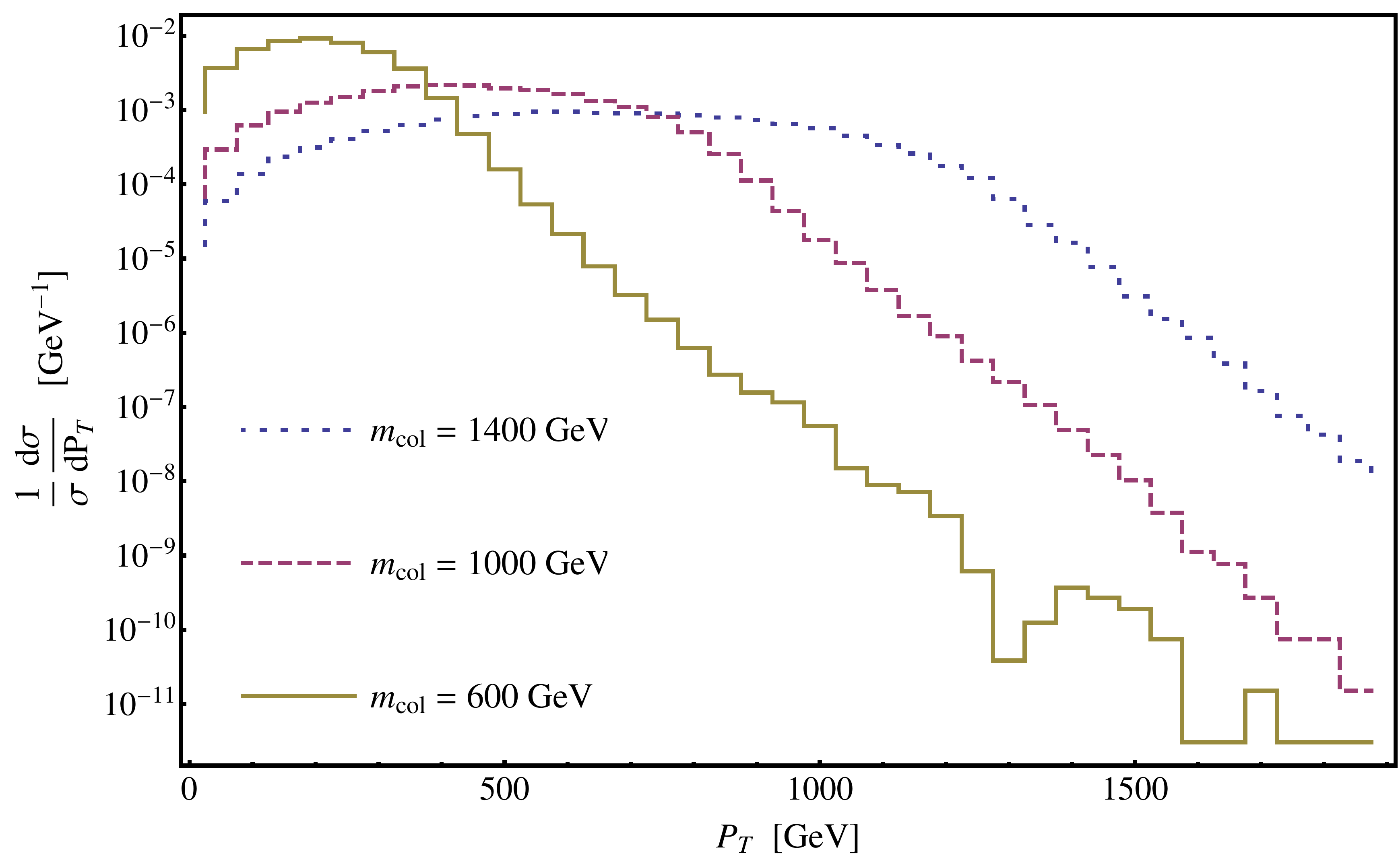}
\caption{  $P_T$ of the hardest jet with new colored state dominating the USR.  Specifically, these colored states are squarks.  The blue dotted line is for $m_\mr{col} =  \mbox{ 1400 GeV}$, the red dotted line is for $m_\mr{col} = 1000 \mbox{ GeV}$ and the yellow solid line is for $m_\mr{col} = 600 \mbox{ GeV}$.  We have fixed $m_p = 300 \mbox{ GeV}$ in all cases.  From Eq.~(\ref{eq:MT2withUSR}), the correction to $M_{T2}$ due to USR is of the form $P_T/m_p$.}
\label{fig:pTcolor}
\end{figure}

As discussed above, the additional radiation shifts all events, including those near the $M_{T2}$ endpoint.  For reference, the radiation distributions for our benchmark models are shown in Figs.~\ref{fig:pTelectroweak} and \ref{fig:pTcolor}.  From Eq.~(\ref{eq:MT2withUSR}), the correction to the $M_{T2}$ endpoint due to USR is of the form $P_T/m_p$.  Hence, the $P_T$ distribution of jets determines how well the parent and child masses can be extracted separately.  From Fig.~\ref{fig:pTelectroweak}, we see that heavier parents lead to harder $P_T$ distributions due to the larger recoil occurring from production of a heavier state.  However, since the correction to $M_{T2}$ goes as $1/m_p$, this enhancement is tempered by the parent mass.  In addition, heavier parents have smaller production cross sections (see Table~\ref{tab:BenchmarkCrosssections}).  Hence, assuming they can be seen above the backgrounds, lower parent mass states give rise to more defined bowls.  The trade-off between background rejection, which is optimized for high masses, and the quality of the $M_{T2}$ bowls, which is optimized for low masses due to the dependence on the ISR, leads to a sweet spot in the range of $\mathcal{O}(200\mbox{ GeV})$ to $\mathcal{O}(500\mbox{ GeV})$, with significant dependence on the spin of the parent.  In the cases with colored states upstream this tension is alleviated since now the $P_T$ distributions are harder and the production cross sections are larger, as in Fig.~\ref{fig:pTcolor}. 

The effects of the radiation on the $M_{T2}$ endpoint are shown by plotting $N(\tilde{m}_c)$ as a function of $\tilde{m}_c$ in Fig.~\ref{fig:exampleBowl}, for 50,000 smuon pair production events with two muons and missing energy, with no background events (also see Fig. \ref{fig:bowlsWithCuts}).  As we will show in the next section, the backgrounds can be very efficiently cut away, and will be insignificant near the $M_{T2}$ endpoint (see Sec. \ref{sec:cuts}).  In what follows, we will present statistical error bars on the DM mass determination using the $M_{T2}$ bowl and will discuss in detail various sources of error and their effect on this analysis.  

\subsection{Statistical Analysis of $M_{T2}$ Bowls}

Contributions to adjacent bins in the $M_{T2}$ bowls from the same events imply that it is inappropriate to use simple $\sqrt{N}$ statistics in computing errors.  Removing one event from a given bin in the distribution can in principle remove one event from \emph{each} bin.  Therefore, we utilized the well-known ``bootstrapping'' method to do the statistical error analysis.  We employed the following method when doing this.  We begin by generating a sample of $\mathcal{O}(100,000)$ signal events (we take $\sqrt{s}= 14 \mbox{ TeV}$).
From those $100,000$ events, we choose a subset of size $N_\mr{events}$, and make 100 independent random selections of $N_\mr{events}$ events from the original data set.  Then for each of these selections we calculate $N(\tilde{m}_c)$ using Eq.~(\ref{eq:bowl}).  This gives us a random sampling of bowls for a given number of events.  Since there is often a degeneracy of minima for each of these random bowls, especially for a low number of events, we take the geometric mean of these multiple minima to give us an average minimum for each bowl.  Note that we do this assuming the theoretical value of $M_{T2}^\mr{max}$ (see Sec.~\ref{sec:VaryMT2Max}).  Finally, we find the mean and standard deviation of these 100 average minima.  To find the standard deviation we used the formula $\sum (x_i-x_\mr{mean})^2/(N-1)$ and checked to confirm that this corresponds to 1-$\sigma$ error for a Gaussian distribution to good approximation.

This method allows for a statistical sampling of the distribution of possible bowls for a given number of events.  We present our results as a function of $N_\mr{events}$, the number of events \emph{before} any cuts are made.  Note that the events which contribute to the bowl have very special kinematics which allow them to go beyond $M_{T2}^\mr{max}$ -- the overwhelming majority of events will not have any bearing on the mass determination.  Hence, cuts designed to remove backgrounds will not cut away these special events which contribute near the minimum of the $M_{T2}$ bowl where the DM mass determination occurs.  This is an expectation we check explicitly in the next section.\footnote{This assumption is not true when $M_{T2}^\mr{max} \approx m_W$ as in the case of, for example, $m_p =100\mbox{ GeV}$ and $m_c = 25 \mbox{ GeV}$.}  Also note that by working with the mean we will systematically underestimate the DM mass due to the asymmetric shape of the bowl.  This asymmetry is due to the shape of the $M_{T2}$ distribution near the endpoint as a function of $\tilde{m}_c$ -- the slope becomes steeper as $\tilde{m}_c$ is taken larger.  The events used for the bowls were generated using the PGS detector simulator so that they do include detector effects which also adds to the consistent underestimates.  As we discuss in Sec.~\ref{sec:DetectorEffects}, detector simulations must be utilized to determine the required correction to account for this off-set.  Further sources of error are discussed below in Sec. \ref{sec:SourcesOfError}.

\begin{figure}[htp]
\centering
\includegraphics[width=0.7\textwidth]{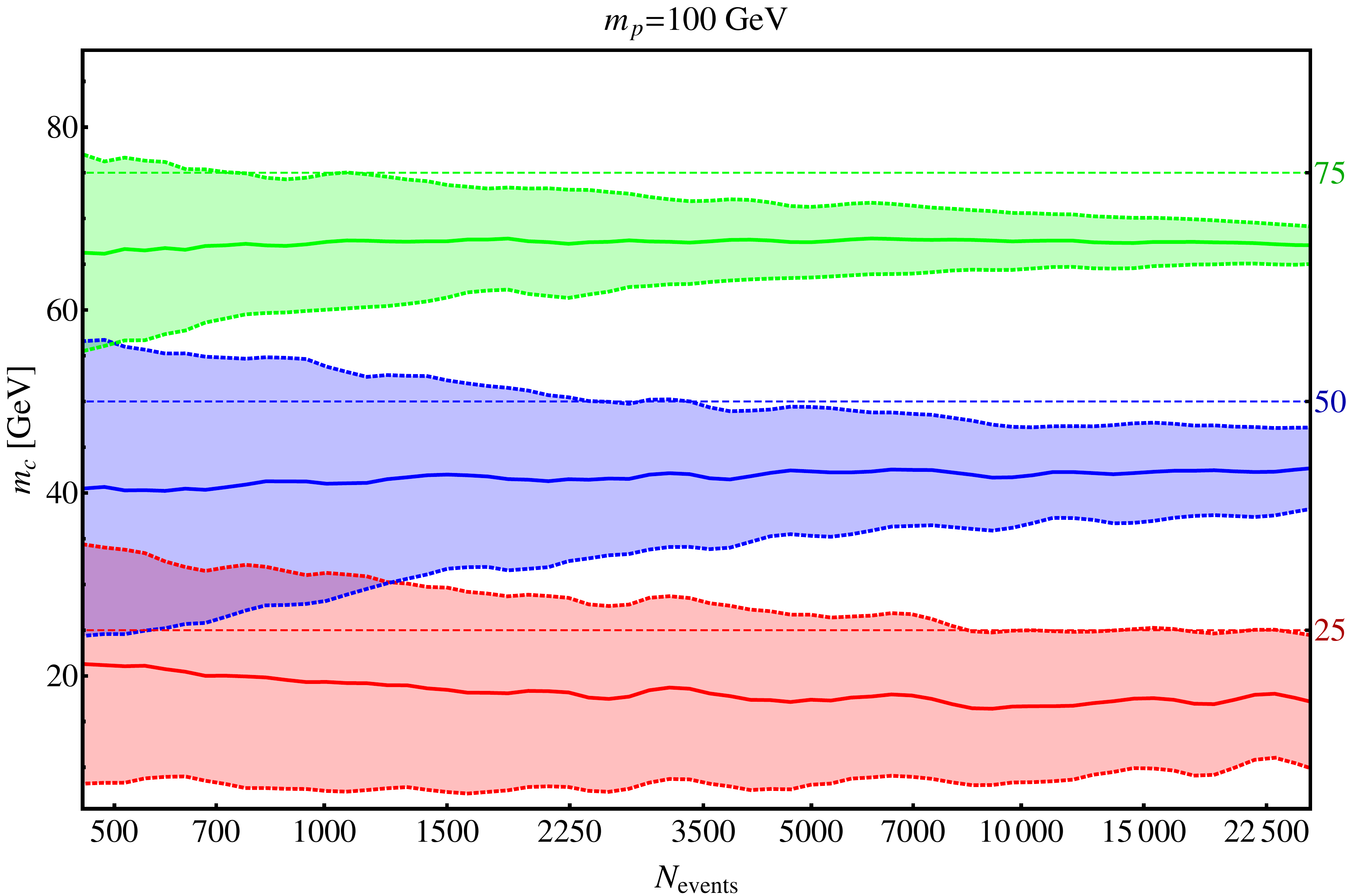}
\caption{Mean and $\pm 1$-$\sigma$ statistical error bars for a DM mass measurement as a function of the number of signal events \emph{before} cuts.  The only source of USR is initial state radiation.  The process we simulated is electroweak smuon production.  The error bars will improve by $\mathcal{O}(1)$ for fermionic parents.  The parent mass is 100 GeV and the child masses are 75 GeV (green), 50 GeV (blue) and 25 GeV (red) from top to bottom.  The dashed lines show the actual child mass.  Note that detector effects have been simulated for the underlying events and that the DM mass measurement systematically undershoots the actual value on account of these effects.}
\label{fig:funnel100}
\end{figure}

\begin{figure}[htp]
\centering
\includegraphics[width=0.7\textwidth]{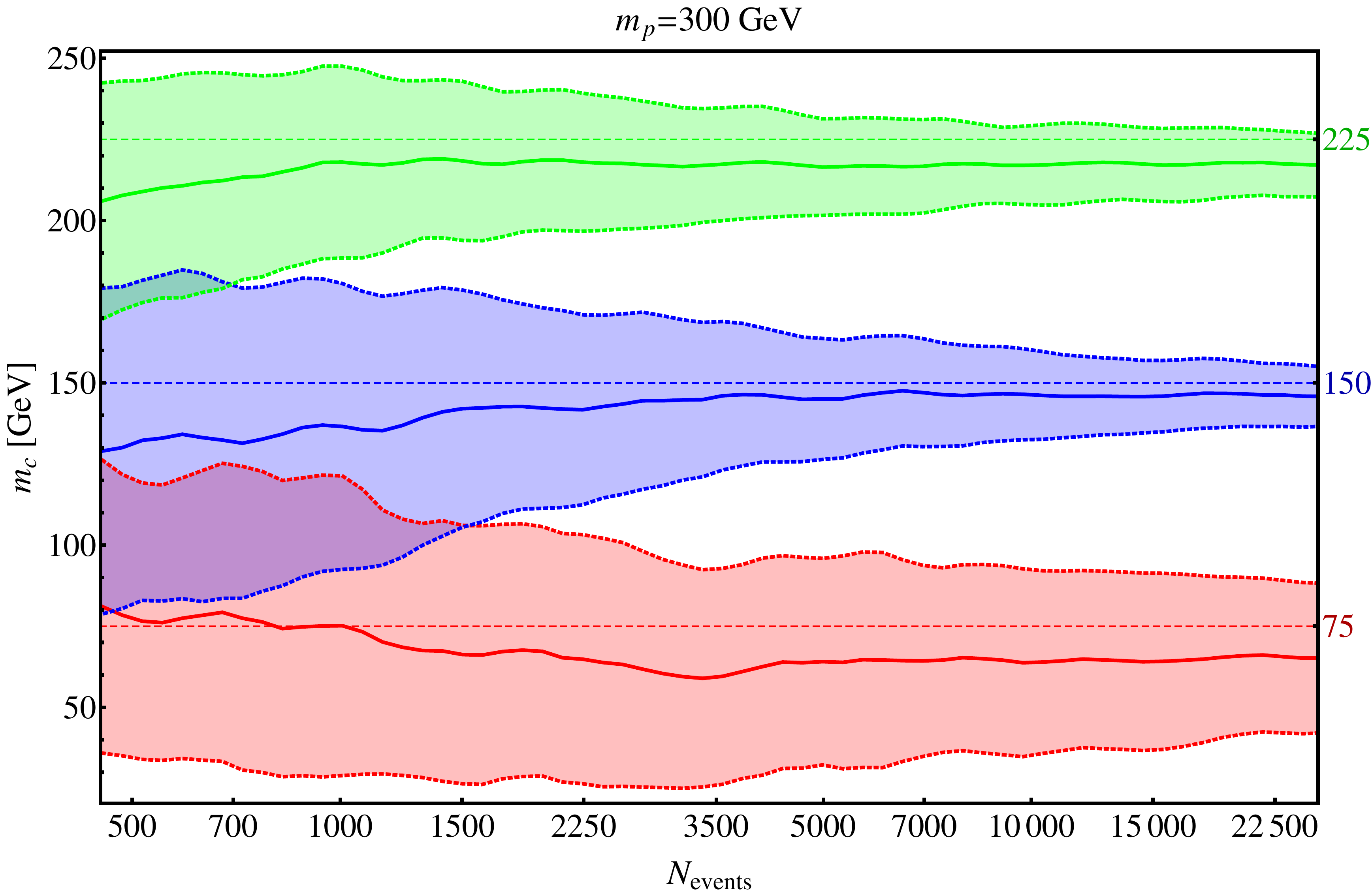}
\caption{Same as Fig.~\ref{fig:funnel100} except that the parent mass is 300 GeV and the child masses are 225 GeV (green), 150 GeV (blue) and 75 GeV (red) from top to bottom.  As explained in the text, we find that cuts designed to eliminate the background will not change these results.}
\label{fig:funnel300}
\end{figure}

\begin{figure}[htp]
\centering
\includegraphics[width=0.7\textwidth]{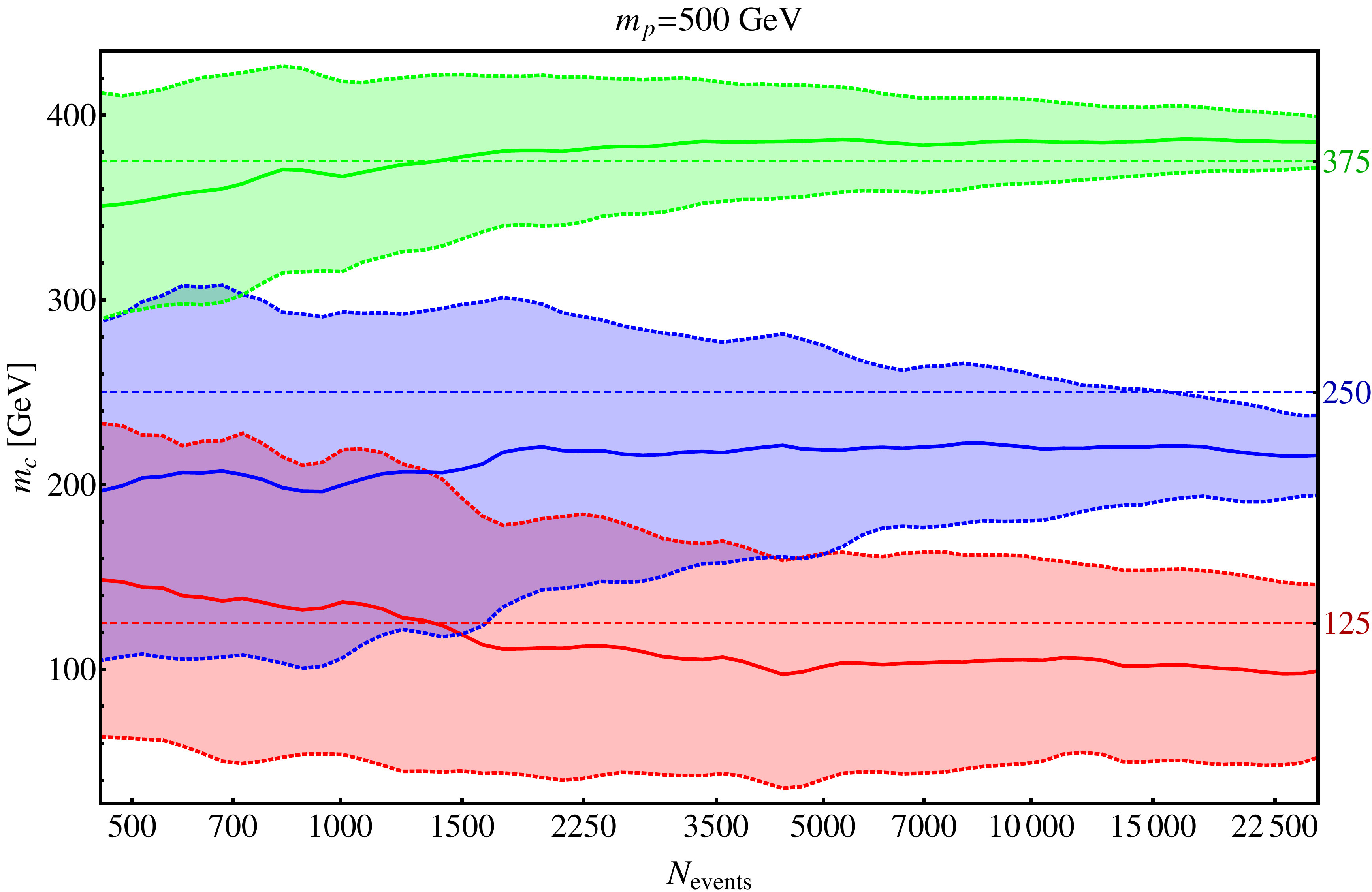}
\caption{Same as Fig.~\ref{fig:funnel100} except that the parent mass is 500 GeV and the child masses are 375 GeV (green), 250 GeV (blue) and 125 GeV (red) from top to bottom.  As explained in the text, we find that cuts designed to eliminate the background will not change these results.}
\label{fig:funnel500}
\end{figure}

\begin{figure}[htp]
\centering
\includegraphics[width=0.7\textwidth]{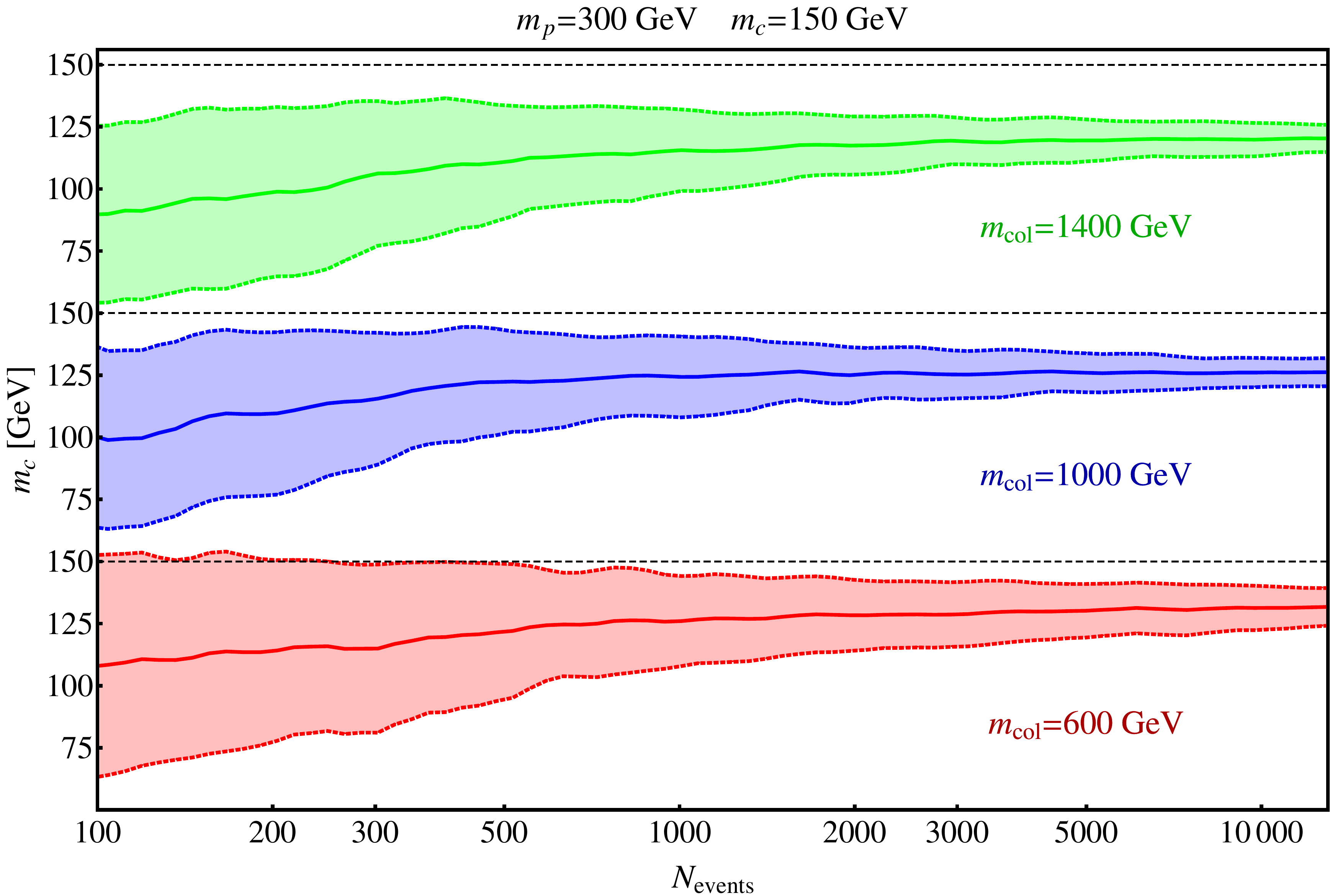}
\caption{Same as Fig.~\ref{fig:funnel300} except that the dominant source of USR is new heavy colored states.  These colored states are squarks which produce chargino parents and jets.  The parent mass is 300 GeV and the child mass is 150 GeV for all three cases.  The mass of the colored objects are 1400 GeV (green), 1000 GeV (blue) and 600 GeV (red) from top to bottom.  As explained in the text, we find that cuts designed to eliminate the background will not change these results.}
\label{fig:funnelsquarks}
\end{figure}

In Figs.~\ref{fig:funnel100} - \ref{fig:funnelsquarks}, we show the statistical error bars for the DM mass determination for a given parent and child mass combination as a function of the number of events before cuts.  Note that for a given child mass, the error bars grow smaller as the DM mass approaches the parent mass, due to the width of the minimum of the bowl.  This occurs because the minimum of the bowl becomes more well-defined as the $M_{T2}$ distribution becomes steeper.  The error bars grow smaller as $N_{events}$ grows larger, but not as quickly as $1/\sqrt{N_\mr{events}}$.  This is because events contribute to multiple bins so that errors from adjacent bins are correlated.  Also notice that error bars in Fig.~\ref{fig:funnelsquarks} are much smaller for a given $N_\mr{events}$ than those in Fig.~\ref{fig:funnel300} for $m_c = 150 \mbox{ GeV}$.  The error bars are also smaller for larger values of $m_\mr{col}$.  This is due to the enhanced $P_T$ of the USR as shown by comparing Figs.~\ref{fig:pTelectroweak} and \ref{fig:pTcolor}.  In what remains we will discuss the various additional errors and will argue to what degree we expect them to degrade the results.

\section{Sources of Error}\label{sec:SourcesOfError}
The results of Figs.~\ref{fig:funnel100} - \ref{fig:funnelsquarks} only incorporate statistical and detector effects.  In this section we argue that the errors we have included in our analysis are a realistic estimate of the precision with which the DM mass can be extracted from simple cascade decays.  We further qualify the additional sources of error below.

\subsection{Detector Effects}\label{sec:DetectorEffects}
\begin{figure}[htp]
\centering
\includegraphics[width=0.7\textwidth]{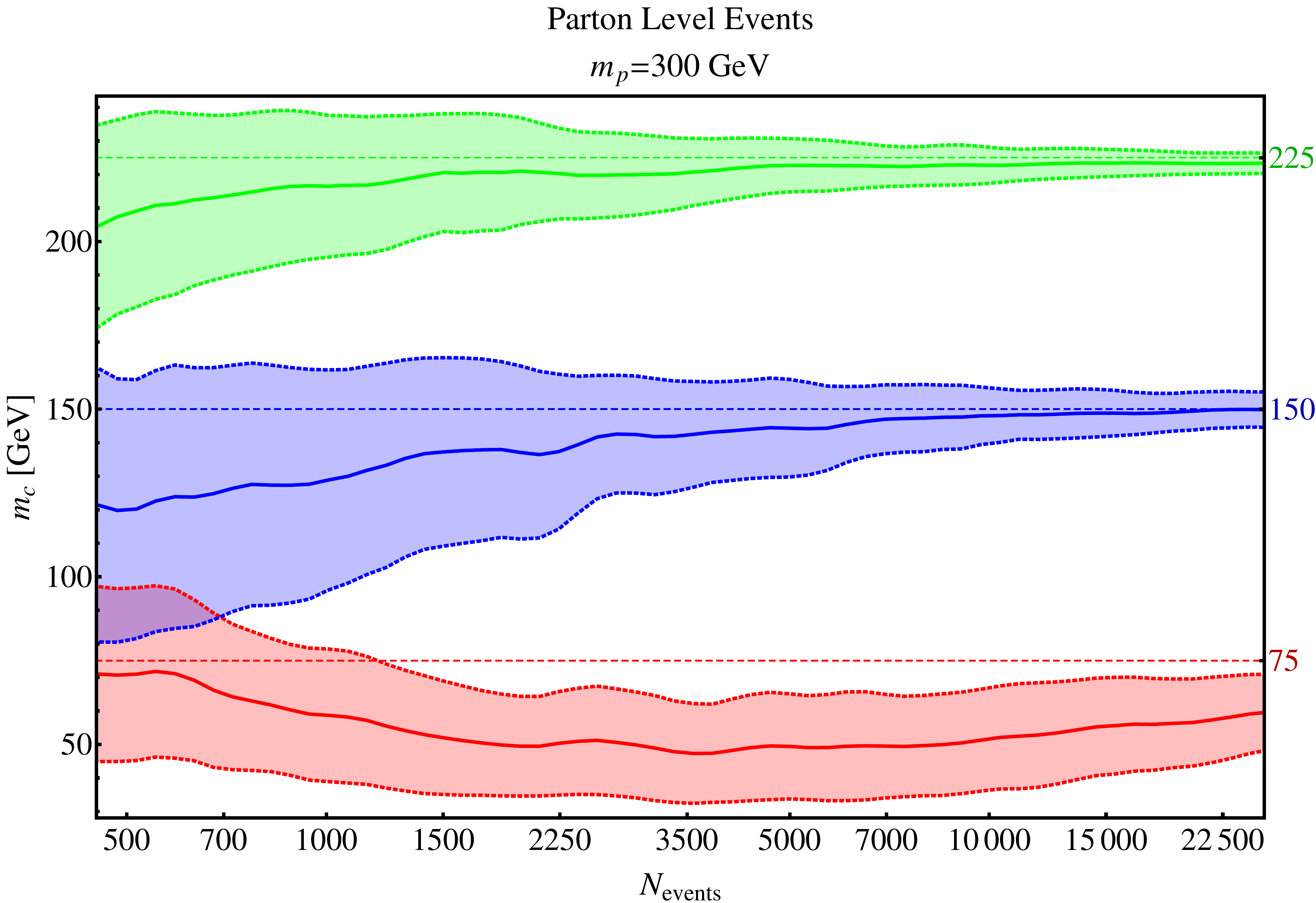}
\caption{Same as Fig.~\ref{fig:funnel300} except that the underlying events are parton level.}
\label{fig:funnelParton}
\end{figure}

With the inclusion of detector effects, the events at the $M_{T2}$ endpoint become smeared out.  This implies that some events which do not have the correct kinematics to make a contribution to the bowl can have $M_{T2} > M_{T2}^\mr{max}$.  This leads to a degradation of the minimum of the bowl.  Since the $M_{T2}$ distribution is steeper for larger test masses, this degradation will tend to contribute to a larger underestimate of the DM mass.  This is the reason for the systematic under-shooting of the DM mass in Figs.~\ref{fig:funnel100} - \ref{fig:funnelsquarks}.  To illustrate this effect we have generated the analog of Fig.~\ref{fig:funnel300} for parton level events as shown in Fig.~\ref{fig:funnelParton}.  Note that the 1-$\sigma$ error bars overlap with the actual DM mass except in the case where $m_c= 75\mbox{ GeV}$ since here the bowl is essentially flat below $\tilde{m}_c \sim 75\mbox{ GeV}$ (see Fig.~\ref{fig:bowlsWithCuts}).  Hence detector simulations would have to correct for this systematic effect in any real DM mass measurement.

After generating bowls using the parton level events, the value $N(\tilde{m}_c)$ (see Eq.~(\ref{eq:bowl})) at the minimum is $\sim 0$.  For the same bowls, but with detector effects, the value $N(\tilde{m}_c = m_c)$ is no longer 0 -- for $\mathcal{O}(100,000)$ events, $N(\tilde{m}_c = m_c) \sim \mathcal{O}(100)$.  Hence one can attempt to clean up the bowl by removing the events from the data sample which contribute at the minimum.  This will increase the steepness of the bowl and might be helpful in minimizing the error since all removed events are guaranteed to be pathological.  However, since this cleaning process does not change the minimum, this will not change the error bars presented above.

\subsection{Background Contamination and Cuts}\label{sec:cuts}
In this section we will argue that a generic set of cuts designed to remove backgrounds will not degrade the minimum of the $M_{T2}$ bowl and hence will not affect our conclusions.  Motivated by the choices taken in \cite{Barr:2005dz}, we have analyzed the following cuts for illustration, which are relevant for di-lepton events with jets and missing energy (\emph{i.e.} slepton pair production):
\begin{enumerate}
\item Require 2 opposite sign, same flavor leptons ($e$ or $\mu$).
\item Hardest lepton: $p_T > 40 \mbox{ GeV}$.
\item Second hardest lepton: $p_T > 30 \mbox{ GeV}$.
\item $p_T^\mr{miss} > 100 \mbox{ GeV}$. 
\item A $Z^0$ veto: the invariant mass of the two leptons, $m_{\ell\ell}$, must not lie in the range \\ $80 \mbox{ GeV} < m_{\ell\ell} < 100 \mbox{ GeV}$.
\item No $b$-tagged jets. 
\end{enumerate}
While cuts should be tailored to the particular model under consideration, these are fairly generic, and will serve to illustrate the point that our results are not significantly degraded by background removal.  We also explored the effect of a cut on $M_{T2}$ by requiring $M_{T2}(\tilde{m}_c = 0) > 100 \mbox{ GeV}$.  These cuts will be very efficient for eliminating standard model (SM) backgrounds, the worst of which is $W^+\,W^-$ plus jets, where the $W^{\pm}$ bosons decay leptonically.  In particular, this di-boson process is dominated by $t\,\bar{t}$ production.  

\begin{figure}[htp]
\centering
\includegraphics[width=0.8\textwidth]{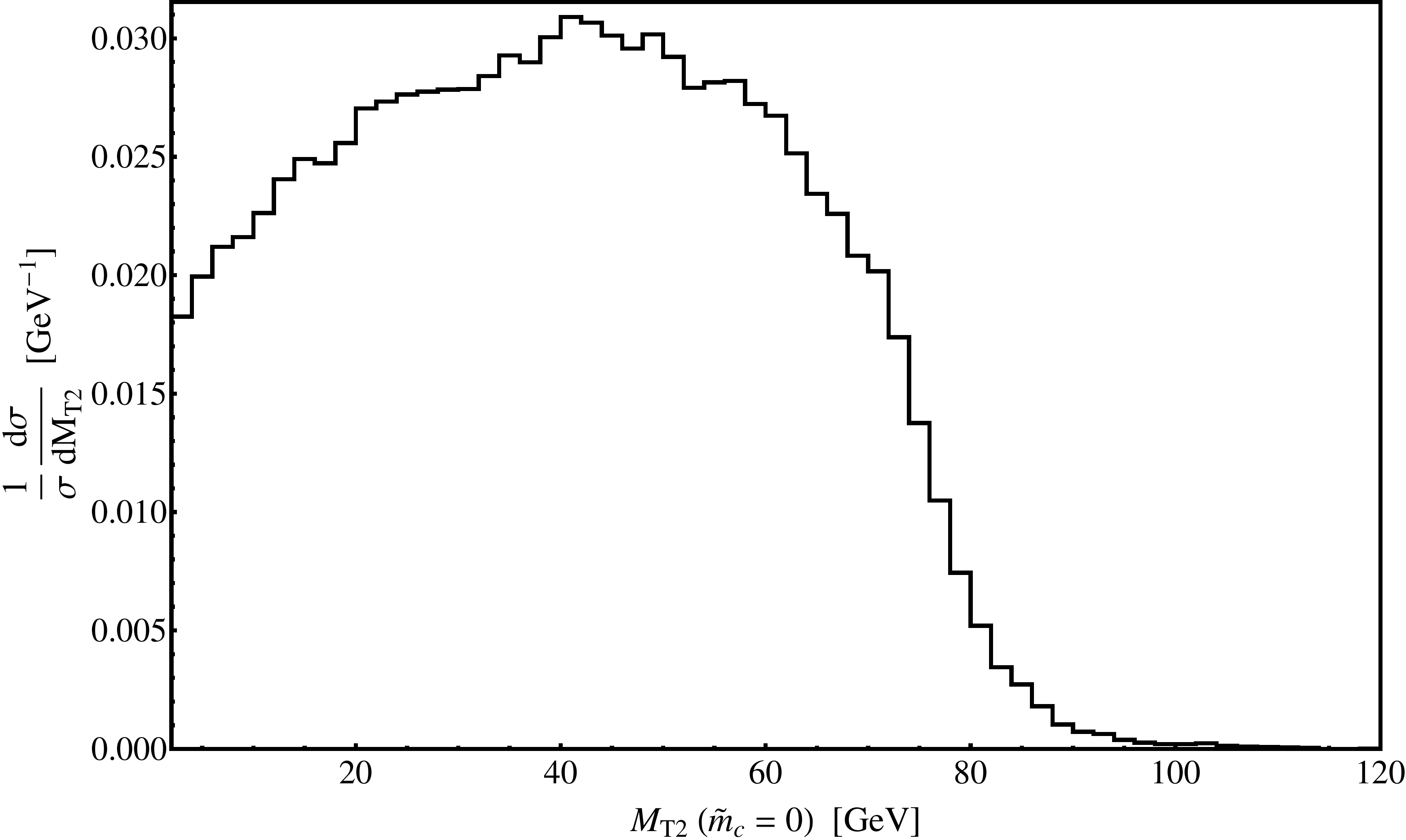}
\caption{$M_{T2}(\tilde{m}_c=0)$ distribution for $t\,\bar{t}$ events where we have treated this as an $n=1$ process where the $b$-jets are USR and the $W^{\pm}$ are the parent particles.  This plot is made before cuts and we have included detector effects.  There is an endpoint at $m_W$ since the child, \emph{i.e.} the neutrino, mass is zero in these events.}
\label{fig:MT2Tops}
\end{figure}

An $M_{T2}$ cut on the $t\,\bar{t}$ background is a powerful discriminator, and in many cases it will have no effect on the DM mass determination.  To see this, first note that the $t\,\bar{t}$ background falls into the same class of $n=1$ processes we have been studying already, with the tops as the colored particles leading to hard USR, the $W^{\pm}$ as parents and the neutrinos as children.  Since the child is a neutrino, $m_c = 0$, and the minimum of the bowl will occur at $\tilde{m}_c = 0$.  Then (neglecting detector effects which will only add a small perturbation) the $t\,\bar{t}$ background will be largely eliminated for an $M_{T2}$ cut of $\mathcal{O}(100\mbox{ GeV})$.  In Fig.~\ref{fig:MT2Tops} we plot this $M_{T2}$ distribution including detector effects.  Clearly, there is an endpoint at $m_W$.  The cross section for $t\,\bar{t}\rightarrow b\,\bar{b}\,\mu^-\,\mu^+\,\nu_{\mu}\,\bar{\nu}_{\mu}$ is 5 pb.  Then starting with a 100,000 event sample, the cuts 1-6 described above reduce this background to $0.065\pm 0.002$ pb.  Then the $M_{T2}(\tilde{m}_c = 0) > 100\mbox{ GeV}$ cut eliminates all remaining events.  In this way, the worst of the SM backgrounds can be easily removed for $m_p \gsim 100 \mbox{ GeV}$.

\begin{figure}[htp]
\centering
\begin{tabular}{cc}
\includegraphics[width=0.5\textwidth]{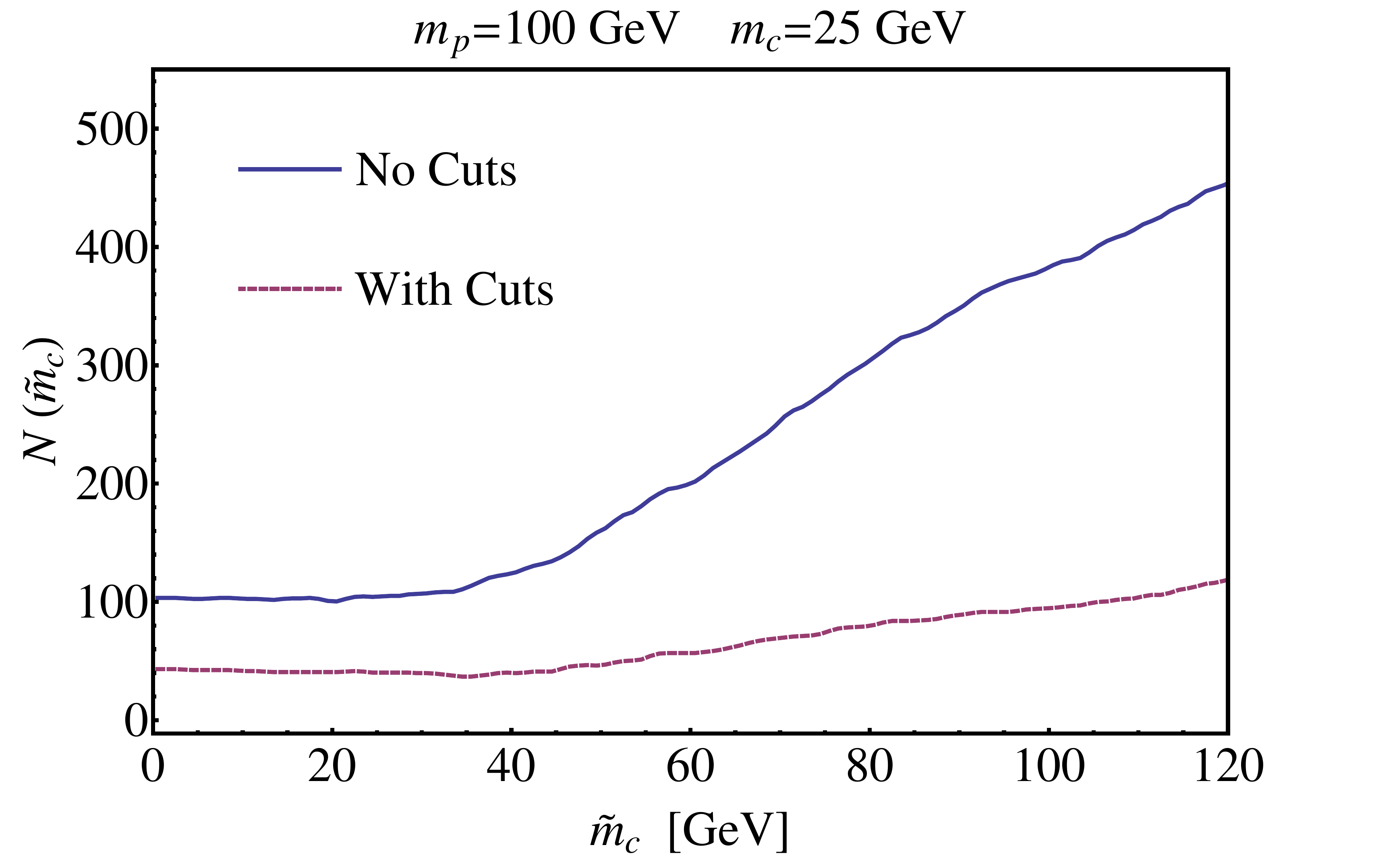} & \includegraphics[width=0.5\textwidth]{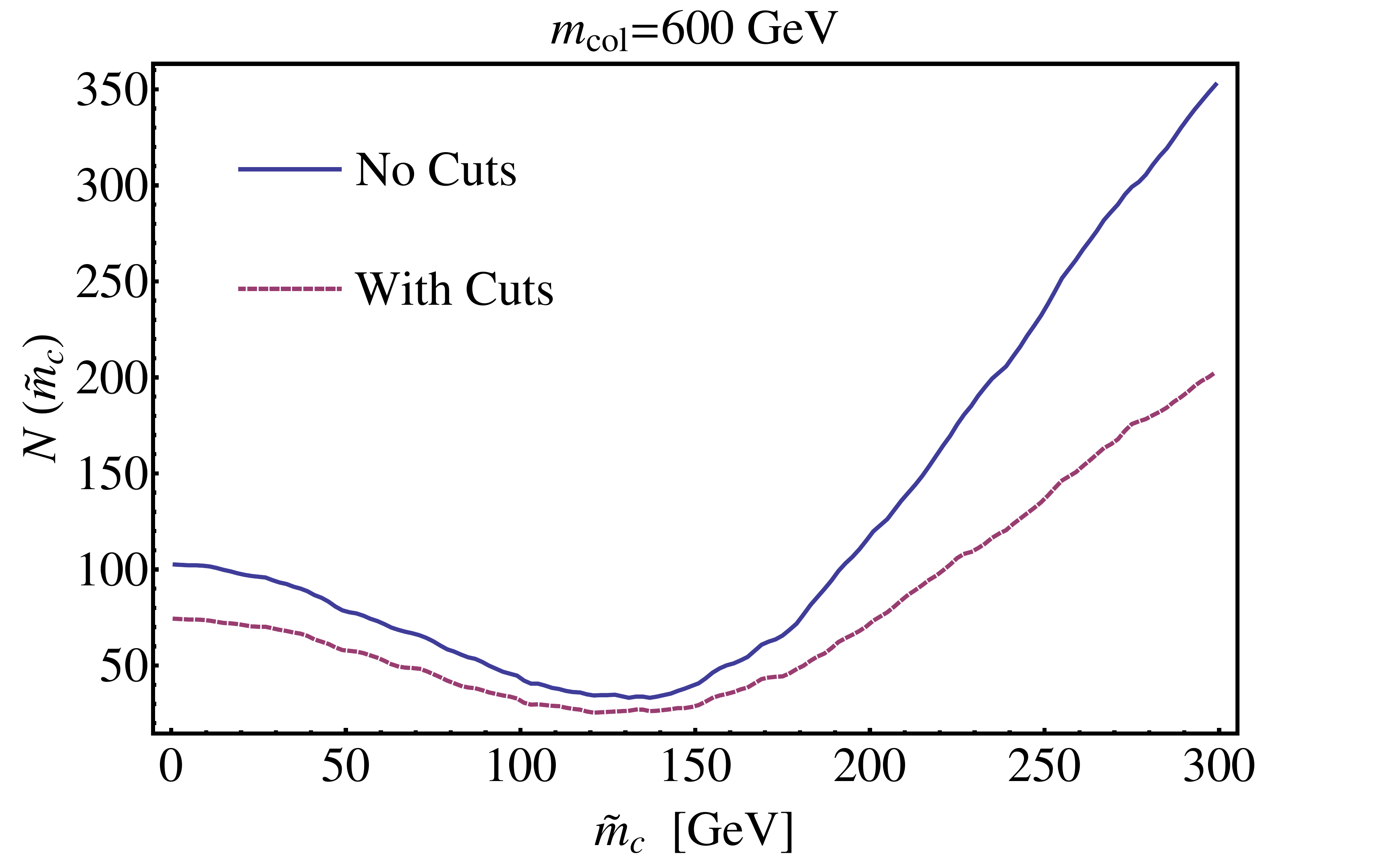}\\
\includegraphics[width=0.5\textwidth]{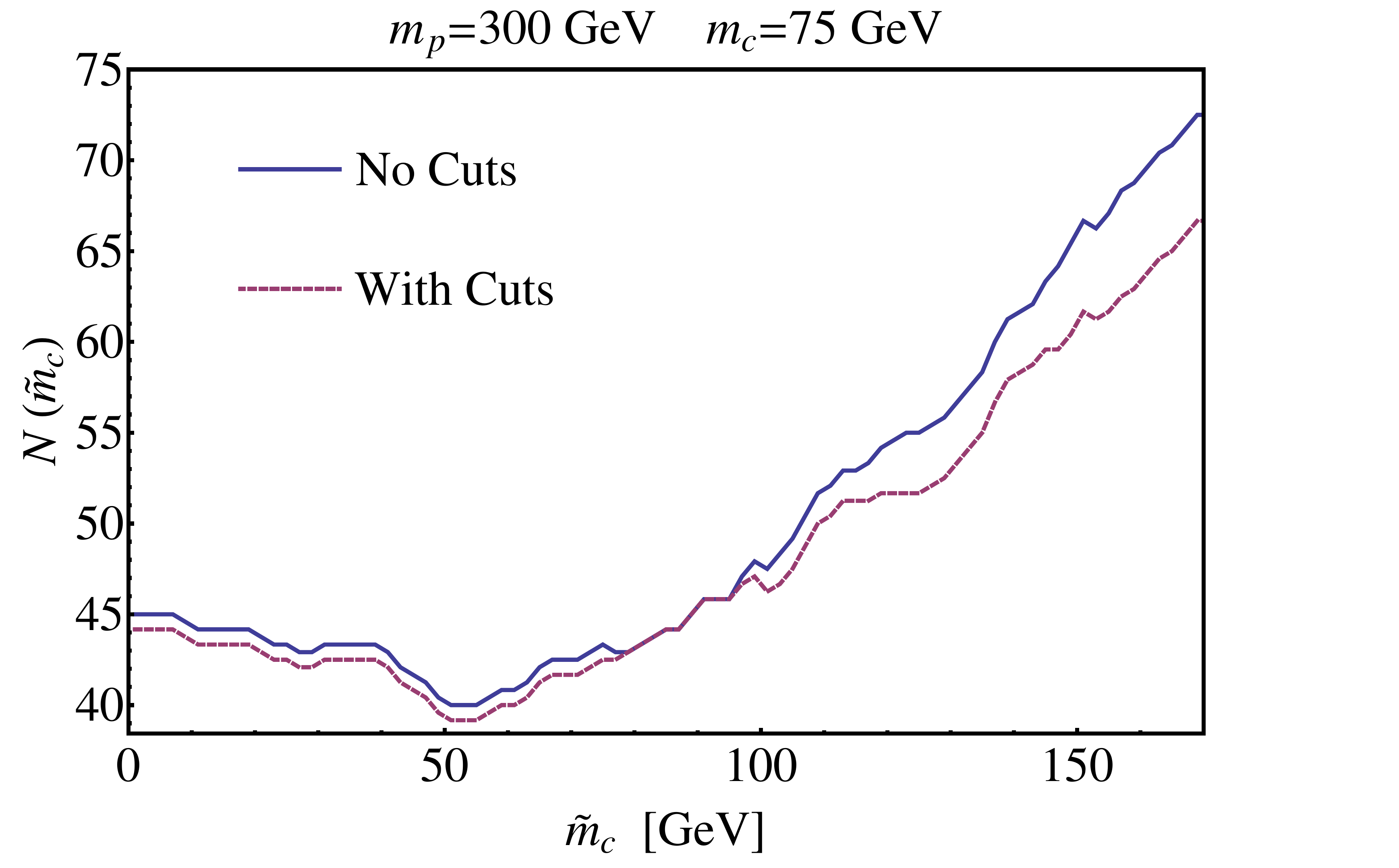} & \includegraphics[width=0.5\textwidth]{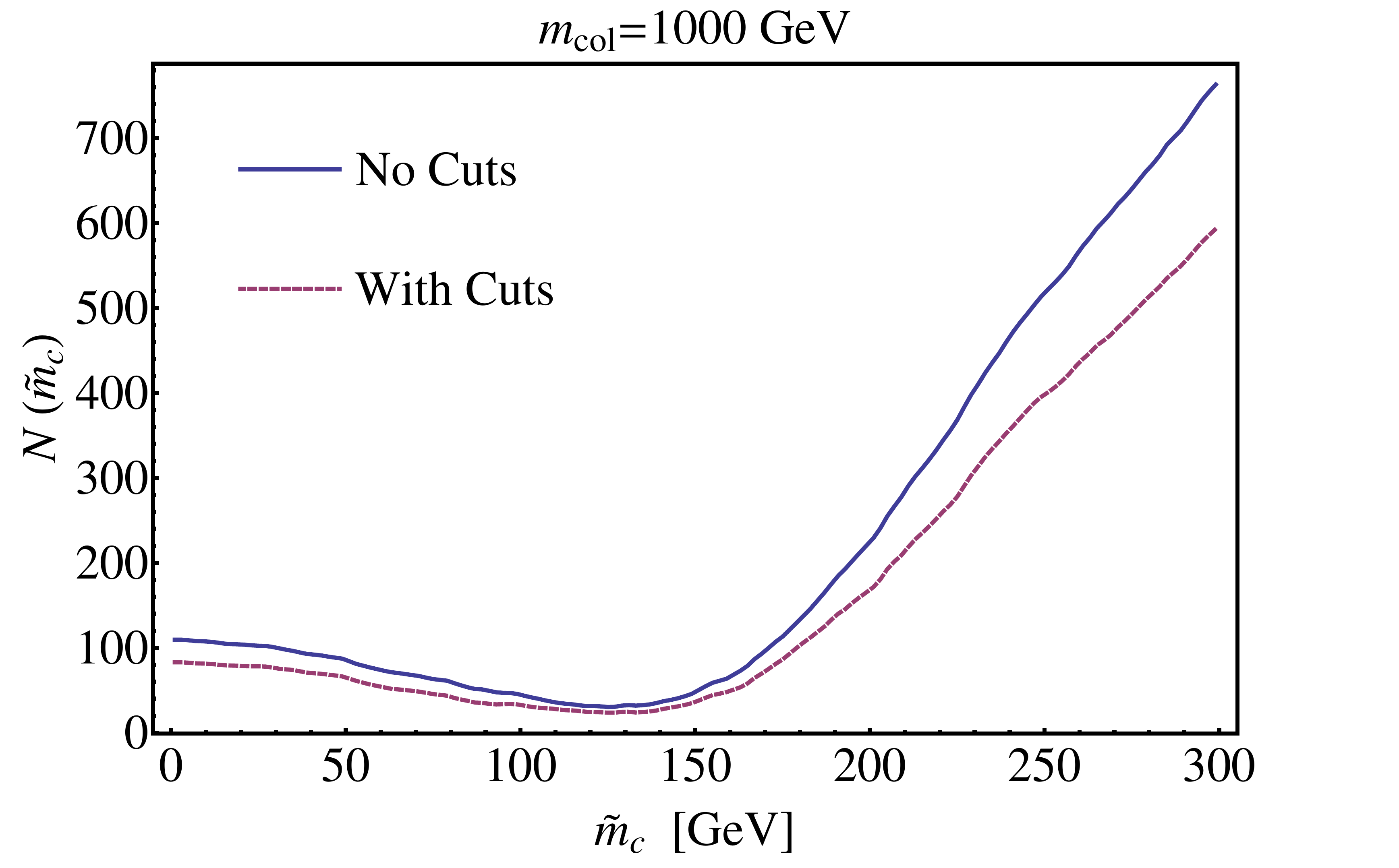}\\
\includegraphics[width=0.5\textwidth]{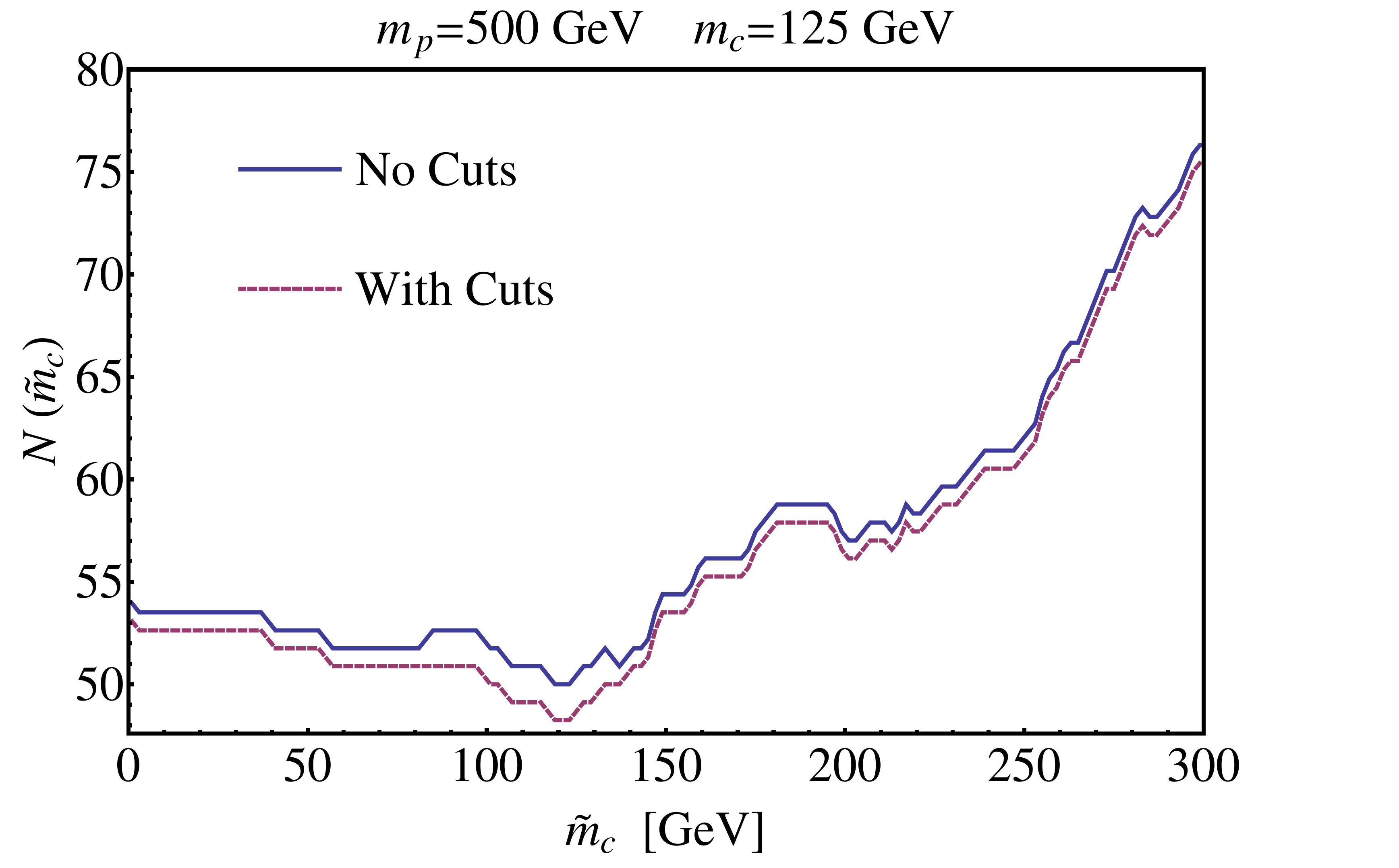} & \includegraphics[width=0.5\textwidth]{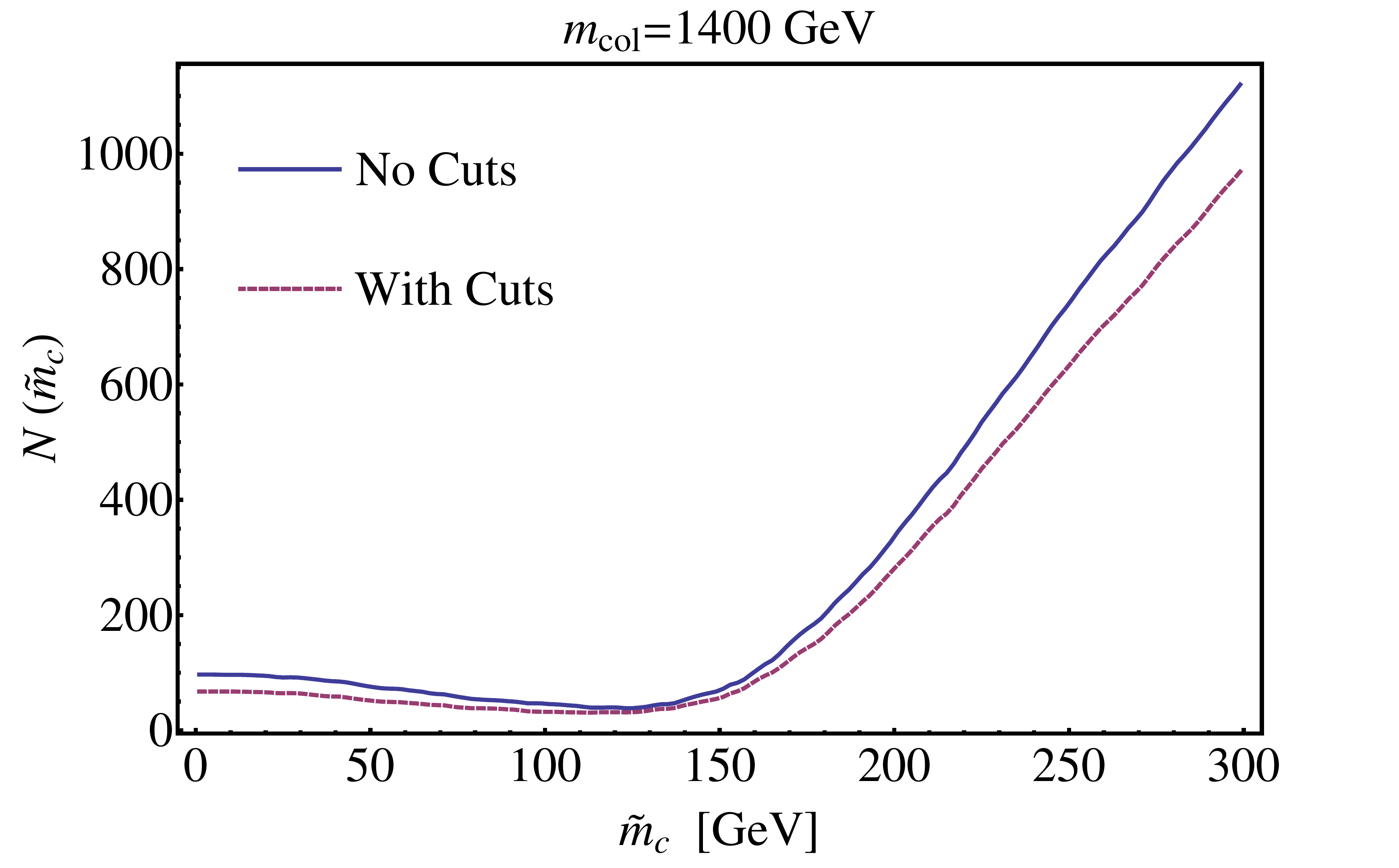}
\end{tabular}
\caption{  $M_{T2}$ bowl for 25,000 (10,000) slepton (squark) pair production events which give jets, two muons and missing energy.  The bowls on the left column only have QCD ISR.  The bowls on the right have additional colored states which dominate the USR, and we have taken $m_p = 300 \mbox{ GeV}$ and $m_c = 150 \mbox{ GeV}$ for these cases.  Note that the cuts preserve the minimum in all cases except $m_p = 100 \mbox{ GeV}$.  Additionally, when one does a cut on $M_{T2}(\tilde{m}_c=0)$, the bowl will be unaffected as long as this cut is taken below $M_{T2}^\mr{max}$ for the bowl in question.}
\label{fig:bowlsWithCuts}
\end{figure}

In Fig.~\ref{fig:bowlsWithCuts} we have plotted a series of $M_{T2}$ bowls before and after this set of cuts to check that the signal in the DM mass determination region of the $M_{T2}$ bowl is not degraded.  For $m_p = 100 \mbox{ GeV}$ there is a significant degradation of the bowl.  However, for this value of $m_p$, there will be tremendous difficulties disentangling the signal from the $W^+\,W^-$ background since they have very similar $M_{T2}$ endpoints.  For the models with heavier parents or with additional colored states producing hard USR, the minimum is maintained for these cuts.  Additionally, the $M_{T2}$ cut has no effect on these plots (excluding the example with $m_p = 100 \mbox{ GeV}$).  We also checked that this statement is robust under variations in the cut parameter choices made above.

\subsection{Variation in $M_{T2}^\mr{max}$}\label{sec:VaryMT2Max}
In generating Figs.~\ref{fig:funnel100}-\ref{fig:funnelsquarks} we assumed that the $M_{T2}^\mr{max}$ endpoint has been measured precisely and matches the theoretical value.  In \cite{bowls}, another $M_{T2}$ based variable, $M_{T2\perp}$, was introduced, which is the projection of $M_{T2}$ along the direction perpendicular to the USR.  They show that the endpoint of this distribution is independent of the USR momentum and identical to $M_{T2}^\mr{max}(\tilde{m}_c,0)$ endpoint.  Hence, even in cases with large USR, it is possible to extract the required input to construct the bowls.  

\begin{figure}[htp]
\centering
\includegraphics[width=0.8\textwidth]{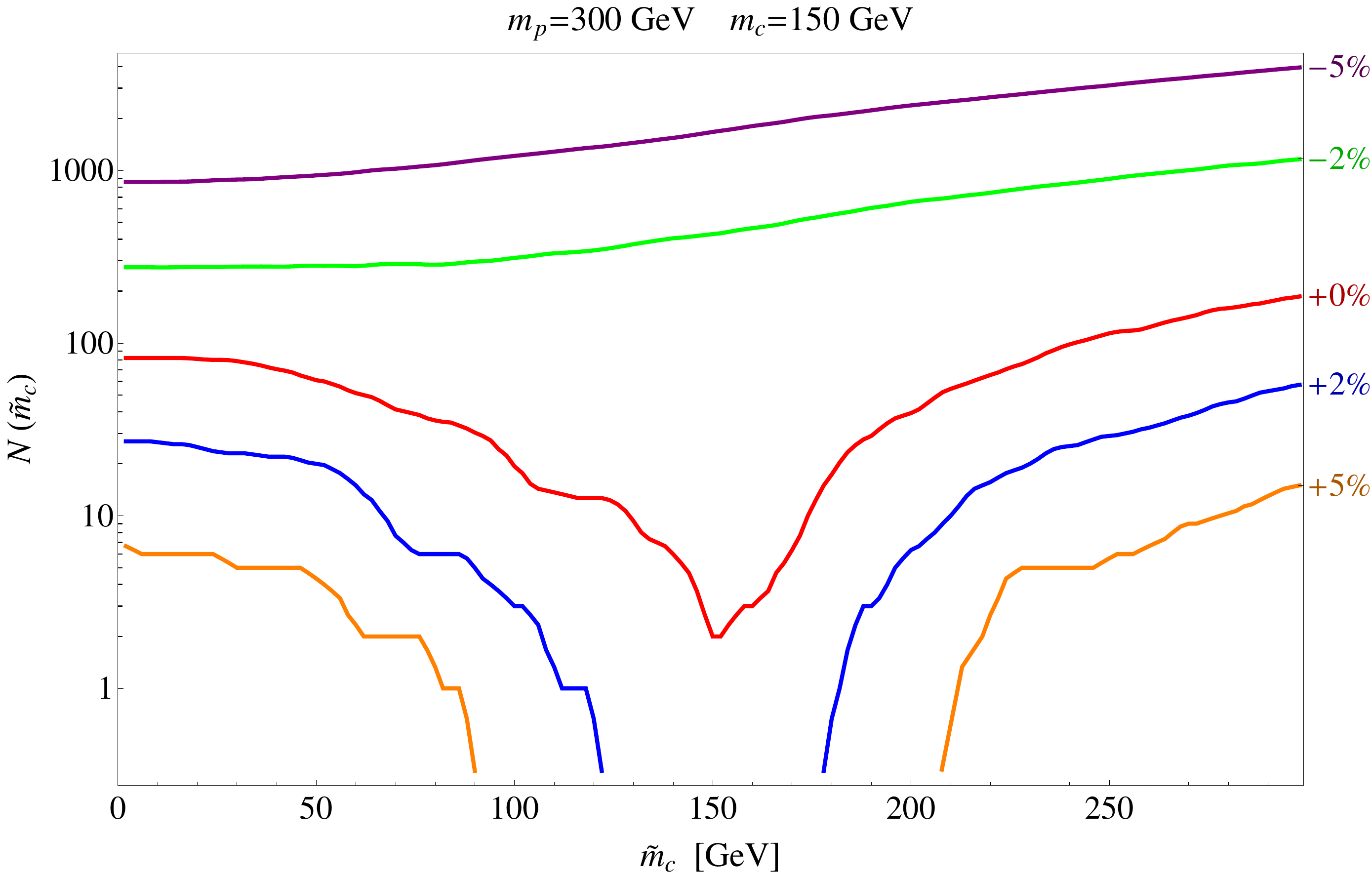}
\caption{  Plot of $M_{T2}$ bowls allowing for variations in the $M_{T2}^\mr{max}$ endpoint of $\pm 5\%$ and $\pm 2\%$.  All bowls are made with 50,000 smuon pair production events before cuts.  For clarity we have not simulated detector effects for these events.}
\label{fig:variationInMT2Max}
\end{figure}

However, the level of accuracy with which $M_{T2}^\mr{max}(\tilde{m}_c,0)$ can be measured depends on detector effects.
For the purposes of illustration, in Fig.~\ref{fig:variationInMT2Max}, we show how the $M_{T2}$ bowl is degraded as one varies the $M_{T2}^\mr{max}$ endpoint by $\pm 2\%$ and $\pm 5\%$ for $m_p = 300 \mbox{ GeV}$ and $m_c = 150 \mbox{ GeV}$.  For variations on the order of $-5\%$ the minimum is shifted by a non-trivial amount and can even disappear in some cases.  For overestimates of $M_{T2}^\mr{max}$ of order $5\%$, the width of the minimum becomes much broader than the statistical error bars presented above.  Therefore, it is crucial to the success of this method that an accurate measurement of the $M_{T2}$ endpoint be made.  On the other hand, the steepness of the bowl around the minimum is maximized for the correct choice of $M_{T2}^\mr{max}$.  By combining this observation with the direct measurement of the endpoint, the accuracy with which $M_{T2}^\mr{max}$ could be determined would be improved.  Other variables, such as that suggested in \cite{} could be used to determine the $M_{T2}$ endpoint.  The accuracy with which this can be done is left for future work, though it can likely be done with high precision due to the larger amount of statistics available than for the bowls.

\section{Discussion and Conclusions}
In this work we studied the possibility of using $n=1$ single stage cascade
decays to measure the DM mass at the LHC.  We have argued, using the particular $M_{T2}$ variant of \cite{bowls}, that if a
signal is observable and backgrounds can be eliminated, it is possible to
make $\mathcal{O}(10\%)$ measurements of the DM mass with
$\mathcal{O}(10,000)$ events before cuts for optimal values of $m_p$ and $m_c$.  We have shown that this requires
a precise determination of $M_{T2}^\mr{max}(\tilde{m}_c,0)$.

In \cite{ayres} the matrix element technique was used to ascertain how well
the neutralino mass could be measured in an $n=1$ squark decay for a
benchmark model with a parent mass of 561 GeV and a child mass of 97 GeV.
Using parton level events so that jet smearing effects, {\em etc.}, are not
considered, they found that with 3000 events before cuts only an upper
limit on the child mass could be determined and with 7500 events a
measurement could be made with an $\mathcal{O}(100\%)$ error
bar.  This can be compared with our Fig.~\ref{fig:funnel500} for the
benchmark $m_p = 500 \mbox{ GeV}$ and $m_c = 125 \mbox{ GeV}$\footnote{Note that since they did not include the effects of ISR, these error bars are a conservative estimate.}.  We find
that with 3000 events we can make an $\mathcal{O}(70\%)$ determination and
for 7500 events error bar goes down to $\mathcal{O}(50\%)$ once the
correction for detector effects is applied as described above in
Sec.~\ref{sec:DetectorEffects}.  Hence, the methods seem to be
competitive, but ultimately a detailed study will be required to determine
which will lead to the best DM mass determination.

Finally, we would like to emphasize the model independence of these
results, even when there are complicated cascade decays.  A large 
class of events can be interpreted as $n=1$ processes with USR.
All that is required is that the only missing energy in event is produced at the
end of the chain as the result of the decay of an on-shell parent, and that the USR be distinguishable from the decay product of the parent.  When this isolation is possible (\emph{e.g.} the two photon plus missing energy signal of some
gauge mediated SUSY breaking models) our results can
be applied up to differences due to detector effects.

\section*{Acknowledgements}
We thank Aaron Pierce, Ran Lu and Liantao Wang for useful comments and discussions.  This work was supported by the DOE Grant \#DE-FG02-95ER40899 (EK) and a String Vacuum Project Graduate Fellow,
funded through NSF grant PHY/0917807 (EK) and by the NSF CAREER Grant NSF-PHY-0743315 (TC). 

\section*{Appendices}

\def\theequation{\Alph{section}.\arabic{equation}}
\begin{appendix}

\setcounter{equation}{0}

\section{Benchmark Models}\label{sec:Models}

The $n = 1$ events studied here are the simplest class of events at the LHC which involve the DM.  Perhaps the most commonly studied of such processes is $p\,p \rightarrow \tilde{q}\, \tilde{q} \rightarrow j\, j\, \tilde{\chi}^0 \,\tilde{\chi}^0$.  In such models, however, one expects there to be higher $n$ processes present as well which will give additional kinematic information.  In this appendix we will outline examples of $n=1$ process with scalar, fermionic and vector parents.  Estimates for the electoweak LHC cross sections for these models are given in Table \ref{tab:BenchmarkCrosssections}.

\subsection{Scalar Parents}
We begin by motivating scalar parents.  Recently, a wave of leptophilic DM models have been proposed to explain measured cosmic ray anomalies.  A non-supersymmetric example, which is additionally motivated by the baryon-DM coincidence, can be constructed by simply extending the SM by two additional fields: a new Higgs doublet, $H'$, and a leptophilic DM state, $X$, interacting via \cite{ADM}
\begin{equation}
{\cal L} = \bar{X}L H' + m_X \bar{X} X.
\end{equation}
When the additional term 
\begin{equation}
\Delta {\cal L} = \lambda (H^\dagger H')^2 + \mbox{h.c.}
\end{equation}
is added to the Lagrangian, where $H$ is the SM Higgs doublet, and the $H'$ is integrated out, the effective operator
\begin{equation}
{\cal L}_\mr{asym}  = \frac{\bar{X}^2 L\,H\, L\,H}{M^4}
\end{equation}
is generated, where $M$ is the effective suppression scale.  This operator transfers the lepton asymmetry to the DM sector, so that the DM density is set by an asymmetry and not thermal freeze-out.  Also note that such leptophilic DM candidates can be viable as an explanation for the observation of an excesses of cosmic ray positrons by the PAMELA experiment \cite{PAMELAADM}.  Although the DM would be asymmetric ({\em i.e.} mostly $\bar{X}$) when its density freezes in, that asymmetry could be erased through Majorana mass terms for $\bar{X}$ and $X$.  Then in the universe today, $X \bar{X} \rightarrow \ell^+ \ell^-$ may give rise to significant cosmic ray positron signals.  

The DM would be created at the collider through the electroweak production of the $H'$, 
\begin{equation}
p\,p \rightarrow H'\, H' \rightarrow X\, \bar{X}\, \ell^+\, \ell^-.
\end{equation}
However, production rates for $p\,p \rightarrow H'\, H' \rightarrow X\, \bar{X}\, \ell^+\, \ell^-$ will be low (see Table \ref{tab:BenchmarkCrosssections}).  While these events could be extracted from the large di-boson background with high luminosity, DM mass determination will be difficult.  

Note that this process is identical to the electroweak pair production of sleptons (see \cite{Barr:2005dz} for a study which determines how feasible it is to find these processes at the LHC),
\begin{equation}
p\,p \rightarrow \tilde{\ell}^+\,\tilde{\ell}^- \rightarrow \ell^+\, \ell^-\,\tilde{\chi}^0\,\tilde{\chi}^0.
\end{equation}

\subsection{Fermionic Parents}
For an example with fermionic parents, we turn to a model which is embedded within the MSSM.  Introduce a superfield DM candidate, $X$, with the quantum numbers of a sterile neutrino.  Then the active sneutrino can mix with scalar partner for $X$, $\tilde{X}$, leading to mixed sneutrino DM.  In \cite{mixedsneutrino}, $\tilde{X}$ has been shown to be a viable DM candidate.  At the LHC, electroweak production can go through 
\begin{equation}
p\,p \rightarrow \tilde{\chi}^+ \tilde{\chi}^- \rightarrow \tilde{X}\, \tilde{X}^*\, \ell^+\, \ell^-,
\end{equation} 
where $\tilde{\chi}^{\pm}$ is a chargino. Since the parent particles are fermions instead of scalars, the production rates are larger (see Table~\ref{tab:BenchmarkCrosssections}). 

With a slight modification, these classes of DM models can be related to the lepton asymmetry.  One can add a new pair of electroweak doublet superfields, $D$ and $\bar{D}$, and a new superpotential term,
\begin{equation}
\Delta \mathcal{W} = m_D \,\bar{D}\, D + \lambda\, \bar{X}\, D\, H_u + y_i\, L_i\, \bar{D}\, \bar{X}+m_X \bar{X} X,
\end{equation}
where $m_D$ is the mass for $D$, $m_X$ is the mass for $X$ and $\lambda$ is a new yukawa coupling.  Integrating out these doublet states results in the lepton number transferring operator
\begin{equation}
\mathcal{W}_\mr{asym} = \frac{\bar{X}^2\, L\,H_u}{M},
\end{equation}
where $M$ is the effective suppression scale.  This operator can be used to generate the relic density.  The production at the collider then goes through the electroweak production of the fermionic $\tilde{D}$: 
\begin{equation}
p\,p \rightarrow \tilde{D}^+\, \tilde{D}^- \rightarrow \tilde{\bar{X}}\, \tilde{\bar{X}}^*\, \ell^+\, \ell^-.
\end{equation}

Production rates in all these fermionic parent models can further be enhanced by embedding the $n = 1$ process into squark decays: 
\begin{equation}
p\,p \rightarrow \tilde{q}\, \tilde{q} \rightarrow \tilde{\chi}^+\, \tilde{\chi}^-\,j\, j \rightarrow \tilde{\nu}\, \tilde{\nu}^* \,\ell^+\, \ell^-\,j\, j
\end{equation}
As described above (see Fig.~\ref{fig:pTcolor}), this will lead to a much harder USR distribution, which in turn will imply better DM mass determination.

\subsection{Vector Parents}
Lastly, we note that within UED models, pair production of vectors gives rise to similar signals.  For example,
\begin{equation}
p\,p \rightarrow W^{(1)+}\, W^{(1)-} \rightarrow \ell^+\, \ell^-\, \nu^{(1)} \bar{\nu}^{(1)},
\end{equation}
where $W^{(1)\pm}$ is a KK $W$-boson, $\nu^{(1)}$ is a KK neutrino is an $n=1$ chain.  This process can similarly be embedded in the decay of new colored states, which gives rise to harder USR:
\begin{equation}
p\,p \rightarrow Q^{(1)}\, \bar{Q}^{(1)} \rightarrow W^{(1)+}\, W^{(1)-} \,j\, j \rightarrow \ell^+\, \ell^-\, \nu^{(1)}\, \bar{\nu}^{(1)}\,j\,j,
\end{equation}
where $Q^{(1)}$ is a KK quark.  Note that if $\nu^{(1)}$ is the DM its mass is restricted to be greater than $\mathcal{O}(50\,\mr{TeV})$ by direct detection experiments \cite{Servant:2002hb}.

\section{Phase Space Dependence on $M_{T2}$}\label{sec:PhaseSpaceAndMT2}

\begin{figure}[htp]
\centering
\includegraphics[width=0.5\textwidth]{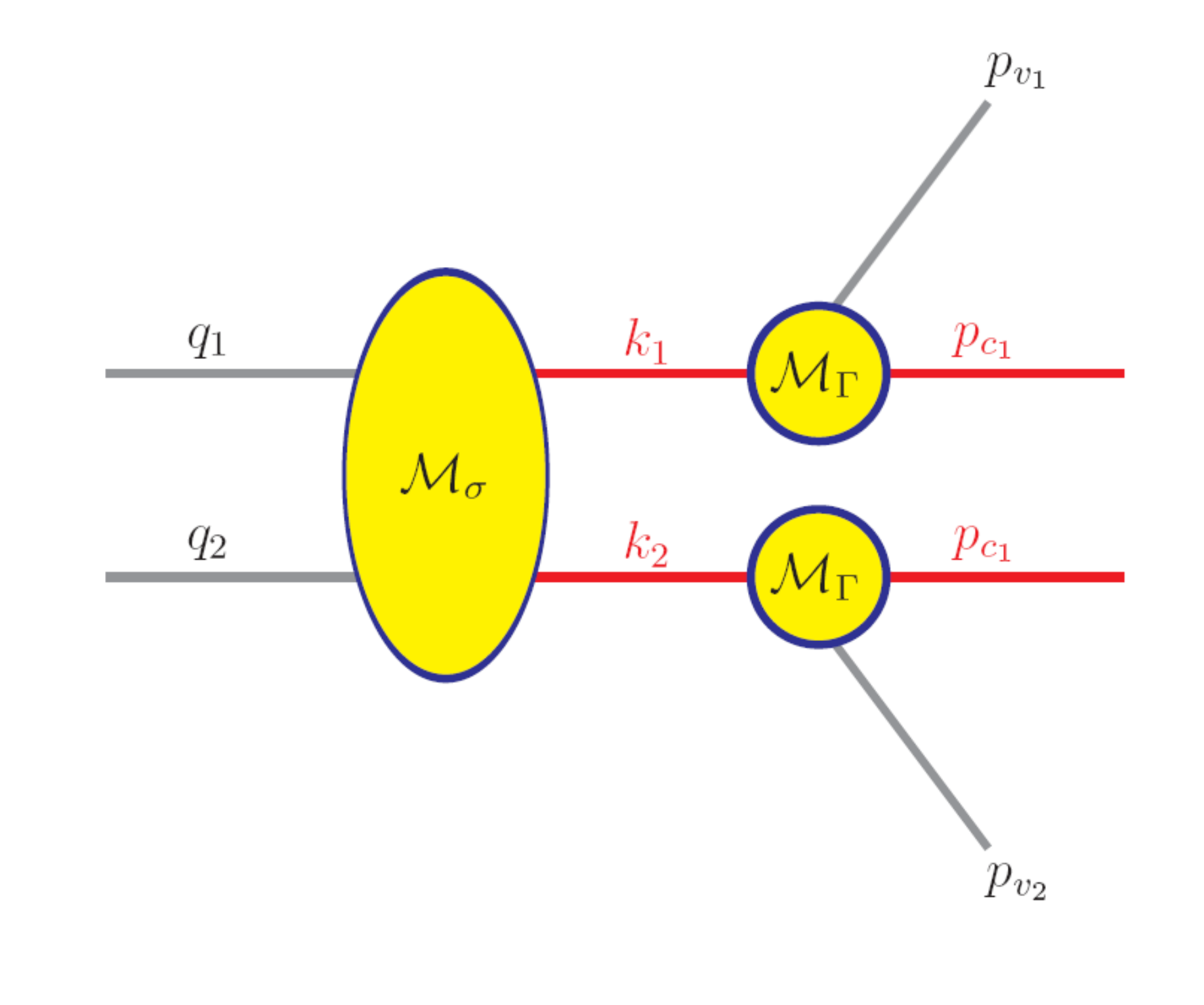}
\caption{Process considered in this section.  The proton momenta are $q_i$, the parent momenta are $k_i$, the visible momenta are $p_{v_i}$ and the child momenta are $p_{c_i}$.}
\label{fig:schematicForMT2Proof}
\end{figure}

To show the phase space dependence on $M_{T2}$ and the overall scale $m_p$, we will assume that the parents are produced on-shell so that the $2 \rightarrow 4$ production in Fig.~\ref{fig:schematicForMT2Proof} can be approximated by the $2 \rightarrow 2$ cross section $\sigma_{2\rightarrow 2}$ and parent particle decay width $\Gamma$.

We begin by simplifying the general $2\rightarrow 2$ differential cross-section and $1\rightarrow 2$ differential decay width. Throughout the calculation we will drop overall constants since they do not contribute to the normalized distributions. The $2\rightarrow 2$ differential cross-section is given by
\begin{equation} 
\mr{d}\sigma_{2\rightarrow2} = \dfrac{1}{4 \left| \vec{q}_1 \right| _\mr{CM}\sqrt{\hat{s}}} \left| \mathcal{M}_\sigma \right|^2 (2\, \pi)^4\, \delta^4(q_1+q_2-k_1-k_2)\, \dfrac{1}{2\, E_1} \dfrac{\mr{d}^3k_1}{(2\,\pi)^3} \dfrac{1}{2\, E_2} \dfrac{\mr{d}^3k_2}{(2\,\pi)^3},    
\end{equation}
where $\sqrt{\hat{s}}$ is the parton center-of-mass (CM) energy and $E_i$ is the energy of the $i^\mr{th}$ parent.  Integrating over $\vec{k}_2$ in the CM frame to eliminate $\delta^3(\vec{q}_1+\vec{q}_2-\vec{k}_1-\vec{k}_2)$  gives
\begin{equation} 
\mr{d} \sigma_{2\rightarrow2} \propto \dfrac{1}{\hat{s}} \left| \mathcal{M}_\sigma \right|^2 \dfrac{1}{E_1\, E_2 }\, \delta(\sqrt{\hat{s}}-E_2 - E_1)\,  \mr{d}^3k_1,    
\end{equation}
where $\vec{k}_2 = -\vec{k}_1$.
Similarly, we simplify the $1\rightarrow 2$ differential decay widths
\begin{equation}
 \mr{d} \Gamma_i = \dfrac{1}{2\, E_i} \left| \mathcal{M}_{\Gamma} \right|^2 (2\, \pi)^4 \delta^4(k_i-p_{v_i}-p_{c_i}) \dfrac{1}{2\, E_{v_i}} \dfrac{\mr{d}^3p_{v_i}}{(2\,\pi)^3} \dfrac{1}{2\, E_{c_i}} \dfrac{\mr{d}^3 p_{c_i}}{(2\,\pi)^3}, 
\end{equation}
where $c$ and $v$ stand for child and visible, respectively, and $i=1,2$. Integrating over $\vec{p}_{c_i}$ to eliminate $\delta^3(\vec{k}_i-\vec{p}_{c_i}-\vec{p}_{v_i})$ gives
\begin{equation}
\mr{d} \Gamma_i \propto \dfrac{1}{ E_i\,E_{c_i}\, E_{v_i}}\left| \mathcal{M}_{\Gamma} \right|^2 \delta(E_i-E_{c_i}-E_{v_i})\, \mr{d}^3p_{v_i},  
\end{equation}
where the $\delta$-function enforces $\vec{p}_{c_i} = \vec{k}_i - \vec{p}_{v_i} $. Since for $1 \rightarrow 2$ decays the summed and squared matrix elements $\left| \mathcal{M}_{\Gamma_i} \right|^2$ are only functions of the masses, they will not contribute to the normalized distributions.  We drop these factors from here forward.

Convolving the differential parent decay width with the differential $2\rightarrow 2$ cross section, and again dropping overall constant factors gives
\begin{eqnarray}
  \mr{d}\sigma& = &\mr{d} \sigma_{2 \rightarrow 2}\,\mr{d} \Gamma_1\, \mr{d} \Gamma_2   \nonumber\\
&\propto& \dfrac{\left| \mathcal{M}_\sigma \right|^2}{\hat{s}} 
\dfrac{\delta(\sqrt{s}-2\, E_1)\, \delta(E_1-E_{c_1}-E_{v_1}) \,\delta(E_2-E_{c_2}-E_{v_2}) }{E_1^4\, E_{c_1}\, E_{v_1}\, E_{c_2}\, E_{v_2}}\, \mr{d}^3k_1 \mr{d}^3p_{v_1} \mr{d}^3p_{v_2}.
\end{eqnarray}
Define $\cos\beta_i$ to be the angle between the visible particle momenta and the parent particle:
\begin{equation}
\vec{k}_i\cdot\vec{p}_{v_i}   \equiv k_i\, p_{v_i}\, \cos\beta_i.
\end{equation}

Rewriting the phase space delta functions so that $\cos\beta_1$, $\cos\beta_2$, and $k_1/m_p$ are the integration variables, the integrand takes on a more revealing form
\begin{eqnarray}
 \mr{d} \sigma_{2\rightarrow2} \,\mr{d} \Gamma_1 \,\mr{d} \Gamma_2 &\propto&  \mr{d}\left(\frac{k_1}{m_p}\right)\,  \mr{d}p_{v_1}\,\mr{d}p_{v_2}\, \mr{d}\Omega_1 \, \mr{d}(\cos\beta_{v_1}) \,\mr{d}(\cos\beta_{v_2})\,\mr{d}\phi_{v_1}\,\mr{d}\phi_{v_2} \nonumber\\ 
&\times& \dfrac{\left| \mathcal{M}_\sigma \right|^2}{\tilde{s}^{5/2}\, \sqrt{\tilde{s}-1}}\,\mathcal{J}(\theta_1,\theta_{v_1},\theta_{v_2}:\theta_1,\,\beta_{v_1},\beta_{v_2})\,\delta \left(\frac{k_1}{m_p} - \sqrt{\tilde{s}-1} \right)\nonumber \\ 
& \times & 
 \delta \left( \cos\beta_1 - \frac{\mu-p_{v_1} \sqrt{\tilde{s}}}{p_{v_1} \sqrt{\tilde{s}-1}} \right) \, \delta\left(\cos\beta_2-\frac{\mu-p_{v_2} \sqrt{\tilde{s}}}{p_{v_2} \sqrt{\tilde{s}-1}} \right),\label{eq:phaseSpaceDependence}
\end{eqnarray}
where $\mathcal{J}(...)$ is the Jacobian for converting from integration over the $\theta$ angles to the $\beta$ angles, which does not depend on any mass parameters, $\mu$ is defined as in the main body of the paper (see Eq.~(\ref{eq:muDef})), and
\begin{equation}
\tilde{s} \equiv \dfrac{\hat{s}}{4\, m_p^2}.
\end{equation}

Hence, Eq.~(\ref{eq:phaseSpaceDependence}) shows that the phase space only depends on the parent and child mass through the two functions $\tilde{s}$ and $\mu$.  Note that when integrating over the parton distribution functions (PDFs), the factor $1/(1-\tilde{s})^2$ will cause the differential cross-section to be dominated by values $\tilde{s} \sim 1$, which corresponds to threshold production of the parent particles. Therefore, for a trivial cross section matrix element, \emph{the differential cross section only depends on the $m_p$ and $m_c$ through the combination $2\,\mu=M_{T2}^\mr{max}(\tilde{m}_c=0)=(m_p^2-m_c^2)/m_p$.} Note that in some cases, $\mathcal{M}_{\sigma}$ will depend explicitly on $m_p$, or masses of particles being exchanged in the corresponding Feynmann diagram, causing slight deviations in the normalized distributions for the same $M_{T2}^\mr{max}$, but different parent mass.  

\begin{figure}[htp]
\centering
\begin{tabular}{cc}
\includegraphics[width=0.5\textwidth]{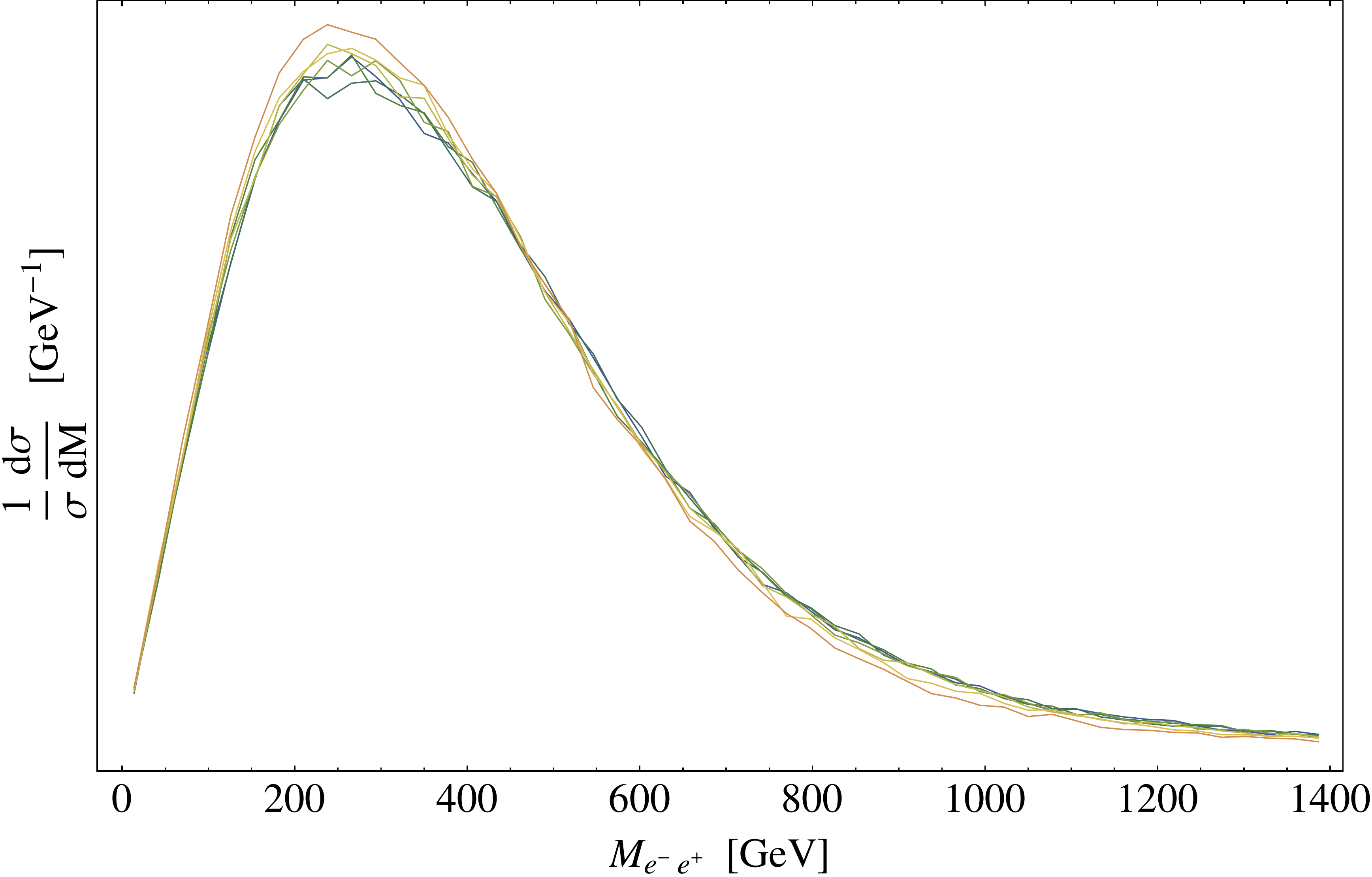} & \includegraphics[width=0.5\textwidth]{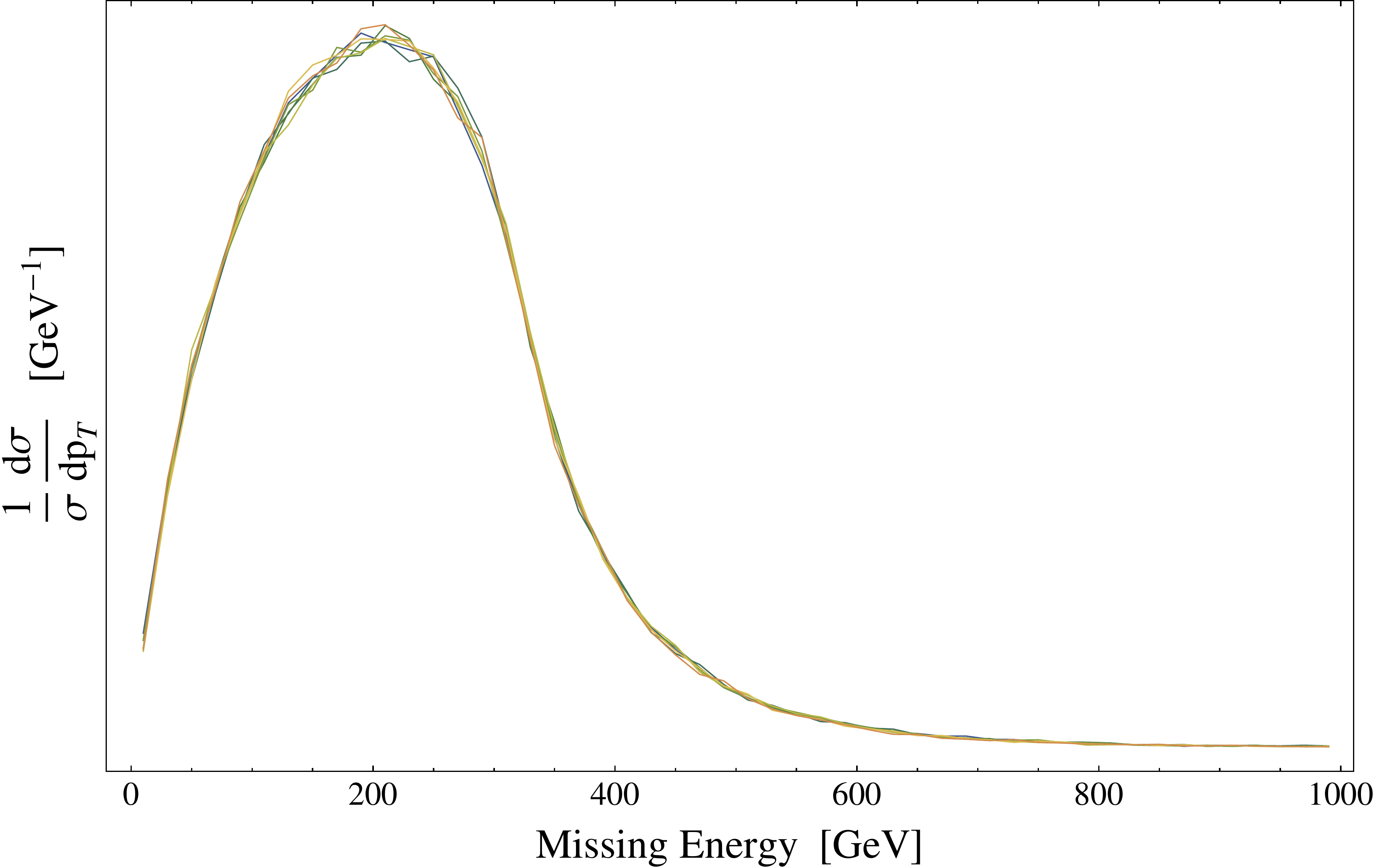}\\
\includegraphics[width=0.5\textwidth]{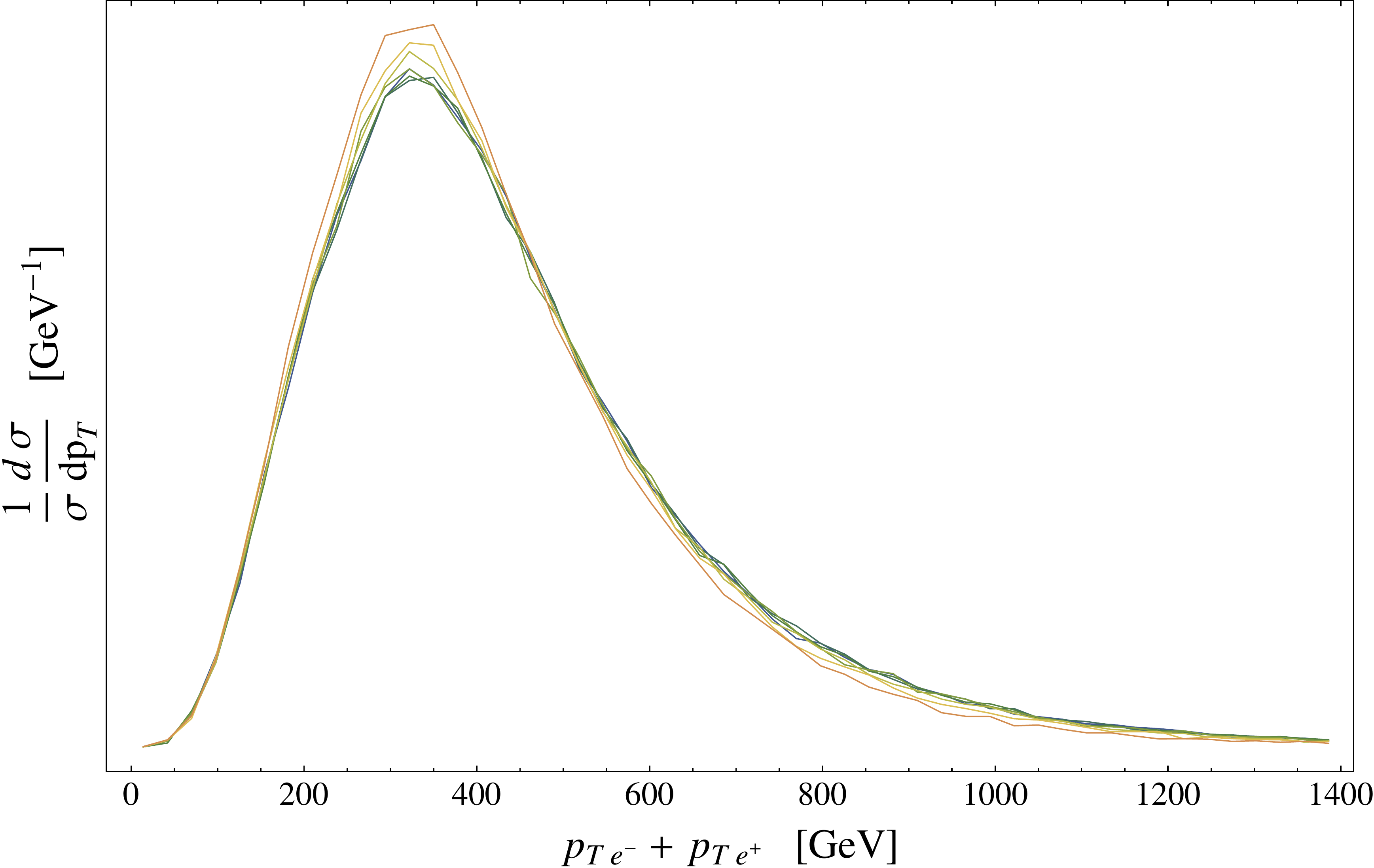} & \includegraphics[width=0.5\textwidth]{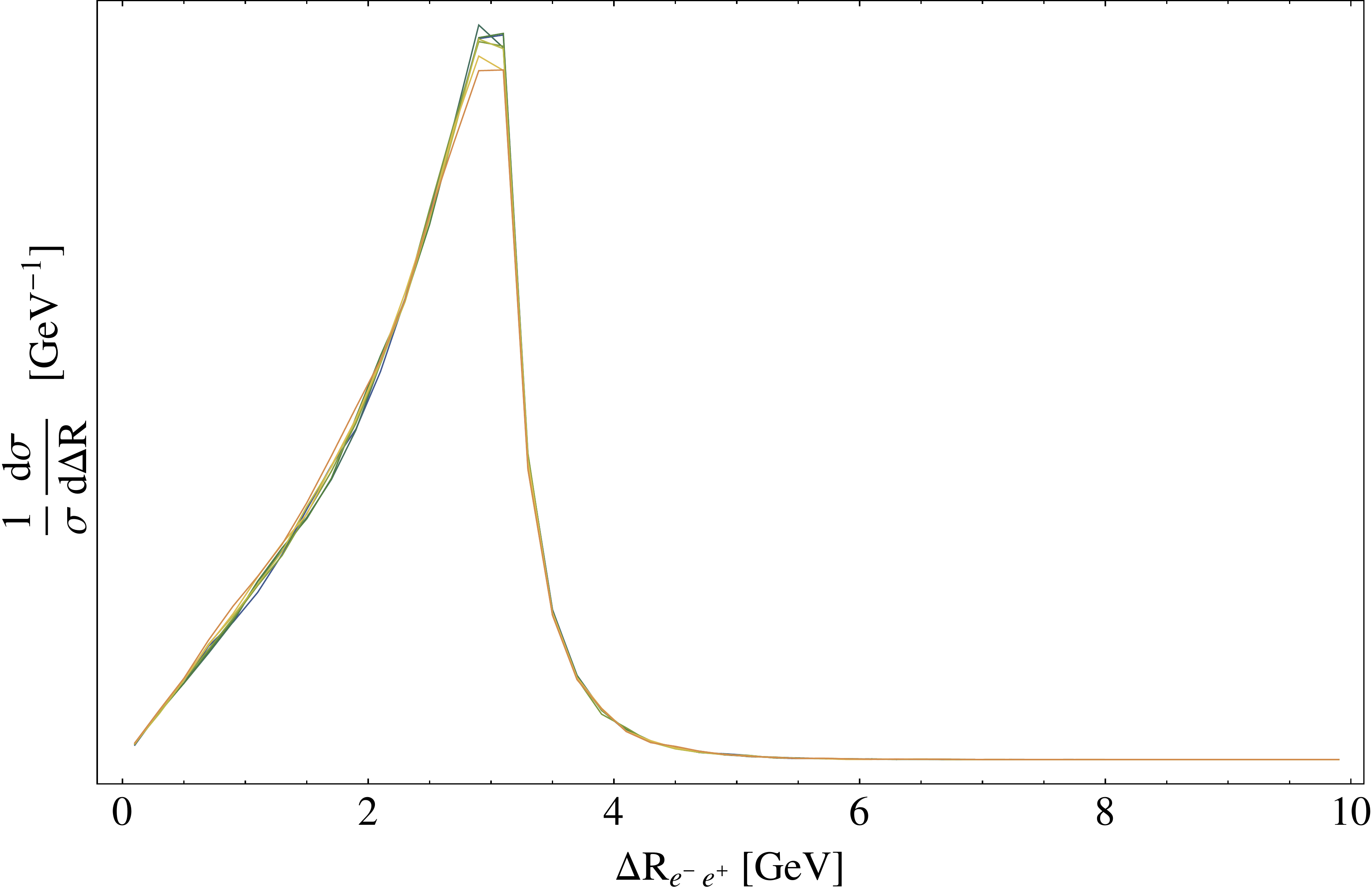} \\
\end{tabular}
\includegraphics[width=1.0\textwidth]{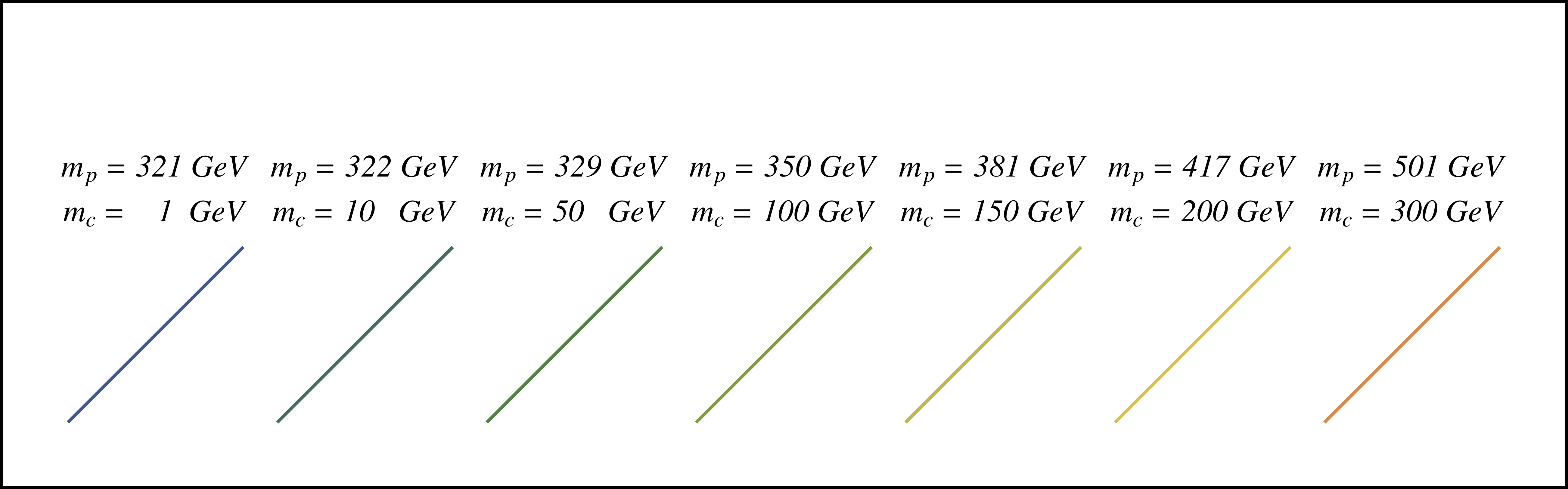}
\label{fig:distributionsSameMT2}
\caption{ Various distributions for points with the same $M_{T2}$ endpoint for $p\,p \rightarrow \tilde{\ell}^+\,\tilde{\ell}^- \rightarrow \ell^+\, \ell^-\,\tilde{\chi}^0\,\tilde{\chi}^0$. As shown in Appendix \ref{sec:PhaseSpaceAndMT2}, the data only depends on $M_{T2}^\mr{max}$ with a slight variation due to $\hat{s}/(4\,m_p^2)$.  The distributions plotted are the invariant mass of the two visible particles (upper left), the total missing transverse energy (upper right), the total transverse momentum of the visible particles (lower left), and $\Delta R = \sqrt{\Delta \eta^2+\Delta \phi^2}$ between the two visible particles (lower right).}
\end{figure}

In Fig.~\ref{fig:distributionsSameMT2}, we plot various normalized distributions with the same $M_{T2}^\mr{max}(\tilde{m}_c=0)$ endpoint.  With the exception of a weak dependence on the parent mass, due to $\hat{s}$ dependence, the distributions look virtually identical.  Note that it is this weak $\hat{s}$ dependence that the matrix element methods seek to capitalize on.

\end{appendix}

\end{document}